\documentstyle[eqsecnum,prd,aps,epsf]{revtex}

\draft 
\begin{document}

\thispagestyle{empty}
{\baselineskip0pt
\leftline{\large\baselineskip16pt\sl\vbox to0pt{\hbox{\it Department of Physics}
               \hbox{\it Kyoto University}\vss}}
\rightline{\large\baselineskip16pt\rm\vbox to20pt{\hbox{KUNS 1518}
            \hbox{\today} 
\vss}}%
}
\vskip1cm
\begin{center}{\large \bf
Final fate of the spherically symmetric collapse of a perfect fluid}
\end{center}
\vskip1cm
\begin{center}
 {\large 
Tomohiro Harada 
\footnote{ Electronic address: harada@tap.scphys.kyoto-u.ac.jp}} \\
{\em Department of Physics,~Kyoto University,} 
{\em Kyoto 606-8502,~Japan}\\
\end{center}

\begin{abstract}
The final fate of the spherically symmetric collapse
of a perfect fluid which follows 
the $\gamma$-law equation of state and 
adiabatic condition is investigated.
Full general relativistic
hydrodynamics is solved numerically using 
a retarded time coordinate, the so-called
observer time coordinate.
Thanks to this coordinate,
the causal structure of the resultant space-time
is automatically constructed.
Then, it is found that 
a globally naked, shell-focusing singularity can occur at the center
from relativistically high-density, 
isentropic, and time symmetric initial data
if $\gamma \alt 1.01$
within the numerical accuracy.
The result is free from the assumption of self-similarity.
The upper limit of $\gamma$ with which a naked singularity
can occur from generic initial data
is consistent with the result of Ori and Piran
based on the assumption of self-similarity.  
\end{abstract}

\section{Introduction}
The singularity theorem~\cite{hp1970}
predicts the existence
of singularity in the generic
gravitational collapse of a massive star.
However it does not state whether or not 
the singularity is covered 
by a horizon.
The naked singularity is considered to be harmful
because it spoils the predictability of physics.
Hence,  
Penrose~\cite{penrose1969,penrose1979} 
presented a cosmic censorship hypothesis.
The weak version says that 
all singularities are hidden in black holes.
The strong version says that
no singularity can be seen by any observer.
We call a singularity censored by the weak version
a globally naked singularity,
and a singularity allowed by the weak version but 
censored by the strong version
a locally naked singularity.
Cosmic censorship has not yet been proved.
In fact, we can easily find that some solutions of Einstein equation
have a naked singularity.
Therefore it is important to find out 
whether those solutions with
a naked singularity
are {\it physically realizable} or not.
For example, we may consider that some energy condition should be
imposed on physical matters.
We should also give regular and generic initial data.
In order to have insight into the physical
reasonableness, it will be helpful to understand
in what case a naked singularity appears.
If a naked singularity is possible in the regime
of classical gravity,
we might catch a glimpse of Planck-scale high-energy
physics or quantum gravity.

The spherically symmetric dust collapse is studied by many
authors because of the existence of an exact solution.
The collapse of a spherically symmetric and homogeneous dust ball 
is described by the Oppenheimer-Snyder solution.
In this solution, a singularity is neither locally nor globally naked.
Therefore any observer cannot see the singularity.
With this solution, 
the usual picture of a black hole as a
final fate of gravitational collapse has been generally accepted.
However, once inhomogeneities of the density and velocity 
distributions are allowed, the above picture does not hold.
In this case, the space-time is given by the 
Lema\^{\i}tre-Tolman-Bondi (LTB)
solution, and it was proved that from very generic initial data
the singularity can be either locally or globally 
naked~\cite{es1979,christodoulou1984,jd1993,sj1996jjs1996}.

The assumption of pressureless matter would not be 
appropriate for high-density matter.
It is obvious that the effect of pressure on the formation of a naked
singularity should be taken into account because 
the formation of a naked singularity
in the LTB solution results in the blow up of the
density.
Ori and Piran~\cite{op198719881990} investigated the collapse of 
a perfect fluid numerically under the assumption of self-similarity.
By this assumption the equation of state of the matter is 
restricted to the form $p=k\rho$.
They showed that a naked singularity forms if $k\alt 0.0105$.
Analytic discussions~\cite{jd19921993} 
based on self-similarity
followed it.
An effort of getting rid of the assumption of self-similarity
was made by Onozawa, Siino, and Watanabe~\cite{osw1994}.
They solved numerically the Misner-Sharp equations~\cite{ms1964} 
from  
regular initial data and searched the formation of an
apparent horizon until the density blows up and
the numerical scheme breaks down.
In fact, their method is not sufficient to
detect naked singularities because the combination of 
the blow up of the density and the absence of an
apparent horizon does not necessarily mean the naked singularity. 

In this paper those difficulties in detection of 
naked singularities are avoided by constructing the null coordinate.
Then, the causal structure of the
space-time can be obtained automatically
by solving the dynamics of the space-time and matter.
Furthermore, by using the ``observer time coordinates,''
the coordinates never cross an event horizon and therefore 
the global nakedness is trivial.

We should note recent progress on the naked singularity
formation in gravitational collapse.
Shapiro and Teukolsky~\cite{st19911992} showed numerically that
a sufficiently prolate (even slowly rotating) 
spheroid of collisionless gas
collapses with the blow up of the curvature invariant
without the apparent horizon formation.
The spherical symmetry of the LTB solution
has been somewhat relaxed.
It was shown that a central, shell-focusing singularity
in nonspherical but quasispherical dust collapse
(i.e., the Szekeres solution~\cite{szekeres1975})
can be either locally or globally naked~\cite{jk1996}.
Recently the stability of the formation of 
the naked singularity in the LTB solution 
against nonspherical linear perturbations 
was investigated numerically, 
and it was suggested that
the Cauchy horizon in the LTB solution is 
stable~\cite{inh1998ihn1998}.  
The restriction to matter has been relaxed
to anisotropic pressure~\cite{dj1994,sw1997,magli1997}.
Harada, Iguchi, and Nakao~\cite{hin1998} showed that
the effect of rotation may induce the naked singularity formation
by studying the collapse of a spherical cloud of 
counter-rotating particles.
Choptuik~\cite{choptuik1993} investigated numerically 
the spherically symmetric collapse
of a scalar field 
and showed that a
zero-mass black hole forms as a critical case for
the black hole formation.

This paper is organized as follows.
In Sec. II, the coordinate systems are presented.
In Sec. III, we discuss an equation of state and initial data.
In Sec. IV, the numerical results are shown.
In Sec. V, we conclude the paper.
We use geometrized units with $c=G=1$ throughout the paper.
We follow Misner, Thorne, and Wheeler's~\cite{mtw1973}
sign conventions of the metric tensor and Riemann tensor.

\section{METHOD}
Here we concentrate on validity of 
the weak cosmic censorship hypothesis, i.e.,
whether or not a singularity can be globally naked.
The singularity is troublesome for numerical relativity
because a numerical scheme
breaks down at the singularity.
If we choose the spacelike hypersurface as a time slice,
we cannot know whether or not the singularity is naked
because it depends on the further evolution of the space-time.
To suggest the nakedness of the singularity, many researchers
have displayed the absence of an apparent horizon.
However the absence of an apparent horizon does not
necessarily mean that the singularity is naked.
In fact, the condition which should be imposed on time slicing
in order to guarantee the singularity avoidance
is not well known (for example, see Ref.~\cite{es1979}).
   
To avoid such difficulties, we adopt the time slicing more suited
to determine the causal properties of the space-time.
For this purpose, we use the outgoing 
null coordinate as a time coordinate.
We determine the scaling of this null coordinate
in accordance with 
the proper time of a distant stationary observer.
This outgoing null coordinate $u$ is called the
``observer time coordinate''~\cite{hm1966}.
This coordinate value corresponds to the time at which a 
distant observer would see the event.
Hence, the observer time coordinates never cross an event horizon.
The limit curve of the time slices $u=\mbox{const}$ in the limit
$u\to\infty$ is, if it exists, an event horizon.
Therefore, if the observer time coordinates hit a singularity,
it turns out to be globally naked.
The procedure to obtain a numerical solution for the 
space-time is as follows.
First, we prepare initial data on a spacelike hypersurface $t=0$.
Then, 
we solve the Misner-Sharp equations~\cite{ms1964} from 
the initial data $t=0$
and store data on the first null ray 
which emanates from the center at $t=0$.
When this ray reaches the stellar surface, we begin to solve
the Hernandez-Misner equations~\cite{hm1966} 
using the stored data on the first null ray 
as initial data $u=0$.
See Ref.~\cite{ms1964,hm1966,vanriper1979,bst1995} for 
basic equations, numerical schemes, and
difference equations.

The code was tested by the collapse of a homogeneous dust ball
and an inhomogeneous dust ball.
A supercritical neutron star collapsed while a subcritical
neutron star did not collapse within many dynamical time scales.
The code was also tested by the Riemann shock tube problem
and intensive explosion described by the Sedov solution.
The conservation of the total mass
is a good indicator of numerical errors.
In all calculations presented in the next section, 
the total mass was conserved within the accuracy of 
$10^{-4}$.
The artificial viscosity term may play a rather subtle role
in the formation of a central singularity.
To avoid such additive difficulties, 
this term was not included basically.
Absence of this term did not spoil the results for 
most calculations because shock wave did not occur
for most cases.
Only for the cases in which the central region expanded,
a shock wave occurred, and thereby the calculation suffered from serious
numerical instabilities, was the artificial viscosity
switched on.
For those cases, the fluid did not collapse, and hence
the center was regular. 
2048 grid zones were prepared in most calculations.

For notational convenience, we give the expression
for the line element in the coordinate systems used here.
In the usual comoving coordinates, the line element 
is written as
\begin{equation}
  ds^2=-e^{2\phi(t,A)}dt^2+e^{\lambda(t,A)}dA^2+R^2(t,A) d\Omega^2,
\end{equation}
while, in the observer time coordinates,
\begin{equation}
  ds^2=-e^{2\psi(u,A)}du^2-2e^{\psi(u,A)}e^{\lambda(u,A)}dudA
  +R^2(u,A) d\Omega^2,
\end{equation}
where
\begin{equation}
  d\Omega^2=d\theta^2+\sin^2\theta d\phi^2,
\end{equation}
and $A$ is chosen to be the rest-mass included 
within the radius $R$ in this article.
In the observer time coordinates, the lapse function $e^{\psi}$
goes to zero when approaching to an event horizon.
The stress-energy tensor for a perfect fluid is given as
\begin{equation}
  T^{\mu\nu}=(\rho+p) u^{\mu}u^{\nu}+p g^{\mu\nu}.
\end{equation}
\section{Initial Data and Equation of State}
Initial data are prepared on the spacelike hypersurface
in order to obtain a clear relation with physical situations.
The initial data are given by the following three arbitrary functions:
\begin{equation}
  \rho_0=\tilde{\rho}_0(R),~~e=\tilde{e}(R),~~U=\tilde{U}(R),
\end{equation}
where $\rho_0$, $e$ are the rest-mass density and 
specific internal energy, respectively.
$U$ is the coordinate velocity defined as
\begin{equation}
  U\equiv e^{-\phi}\frac{\partial R}{\partial t}
  =e^{-\psi}\frac{\partial R}{\partial u}.
\end{equation}
The total energy density is given by 
\begin{equation}
  \rho=\rho_0 (1+e).
\end{equation}
We choose the density distribution as
\begin{eqnarray}
  \tilde{\rho}_0(R)&=&\cases{
    \tilde{\rho}_{0c}\left[1-\left(\frac{R}{R_s}\right)^2\right]
    & ($0\le R\le R_s$), \cr
    0 & ($R_s <R$) \cr}.
\end{eqnarray}
The distribution of the specific 
internal energy and velocity is set as
\begin{eqnarray}
  \tilde{e}(R)&=&\tilde{e}_{c} \left(\frac{\tilde{\rho}_0}
    {\tilde{\rho}_{0c}}\right)^{\gamma-1},
  \label{eq:inite} \\
  \tilde{U}(R)&=&0,
\end{eqnarray}
where $\tilde{\rho}_{0c}\equiv\tilde{\rho}_0(R=0)$.
We use the following $\gamma$-law equation of state:
\begin{equation}
  p=(\gamma-1)e\rho_0.
  \label{eq:eos}
\end{equation}
The combination of the initial data (\ref{eq:inite}),
equation of state (\ref{eq:eos}), and
adiabatic condition guarantees that pressure is in proportional
to $\rho_0^{\gamma}$, i.e.,
\begin{equation}
  p=K\rho_0^{\gamma},
\end{equation}
where $K$ is constant all over the star.
In this case, the initial 
distribution of the specific internal energy is
parametrized only 
by the central specific internal energy $\tilde{e}_{c}$.
If we take the extremely relativistic limit ($e\gg1$), 
the above equation of state becomes
\begin{equation}
  p=(\gamma-1)\rho,
  \label{eq:opeos}
\end{equation}
which is the equation of state used by Ref.~\cite{op198719881990}.

\section{Results}
In determining the final fate of collapse,
here we adopt the following criteria.
If the ratio of the rest-mass density of the innermost 
grid zone to that of the next grid zone excesses 2,
we call it a central `singularity'
and stop the code.
If the lapse function $e^{\psi}$ in the Hernandez-Misner code
decreases to less than $10^{-3}$,
we call it an `event horizon'.
If a singularity occurs before an event horizon is detected,
we call it a `naked singularity.'
The result is not so sensitive to the choice of the thresholds.
Note that the blow up of the rest-mass 
density inevitably results in
the blow up of the scalar $R^{\alpha}_{\beta}R^{\beta}_{\alpha}
=64\pi^2(\rho^2+3p^2)$.
The scalar curvature $R=8\pi(\rho-3p)$ also blows up if $p\ne (1/3) \rho$.

Here, models for three values 
of $\tilde{e}_{c}$, $10^2$, $1$, and $10^{-2}$, were calculated.
If $e \agt 1$, then the fluid is relativistic,
while, if $e \ll 1$, then the fluid is not relativistic.
The results are summarized in Figs. \ref{fg:nsslmshm}--\ref{fg:stcdhm} 
and Tables \ref{tb:hr}--\ref{tb:nr}.

\subsection{naked singularity}
\begin{figure}
      \centerline{\epsfysize 6cm \epsfxsize 9cm \epsfbox{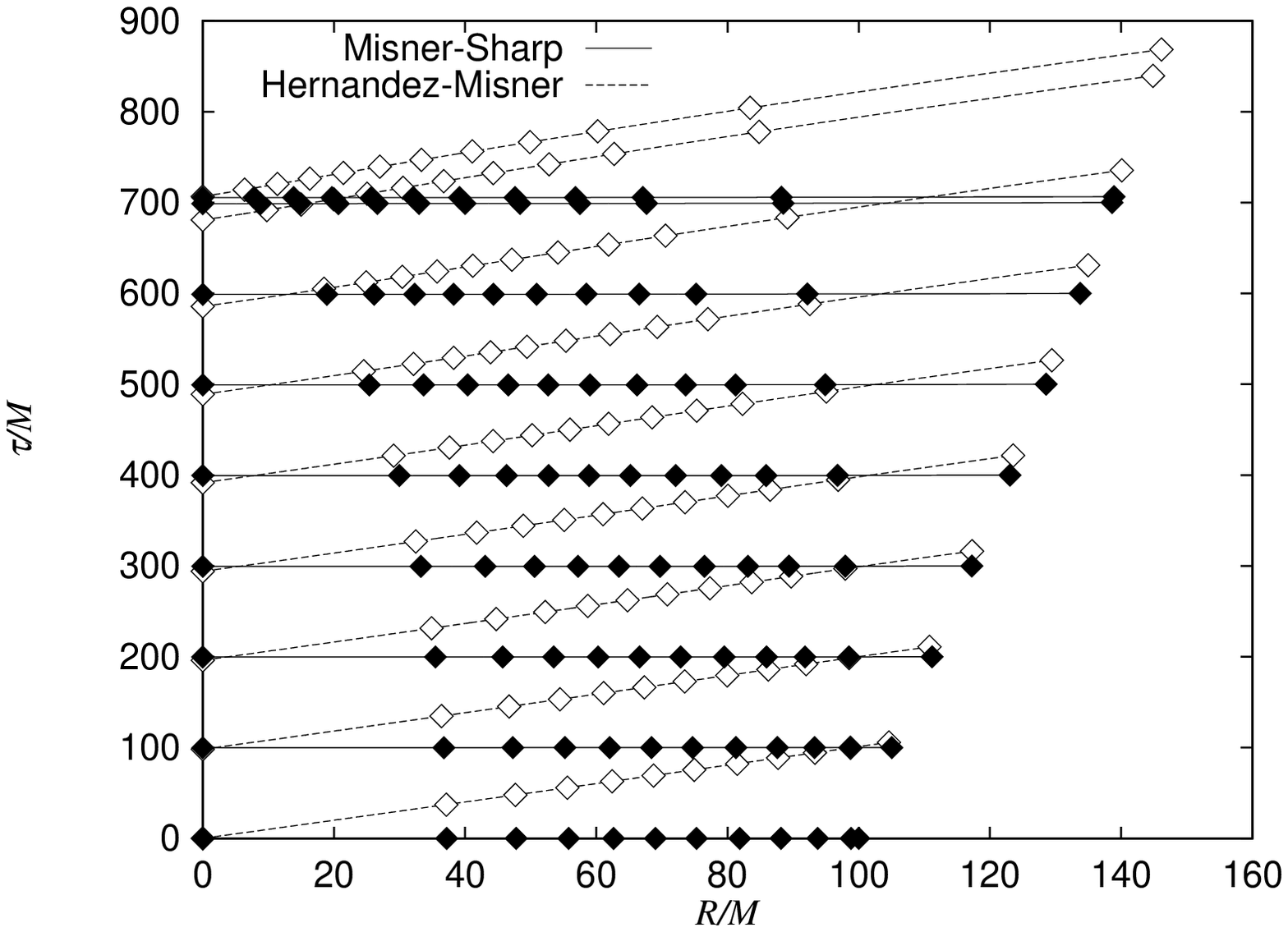}}
      \caption{}
      \label{fg:nsslmshm}
      \vspace{1cm}
      \centerline{\epsfysize 6cm \epsfxsize 9cm \epsfbox{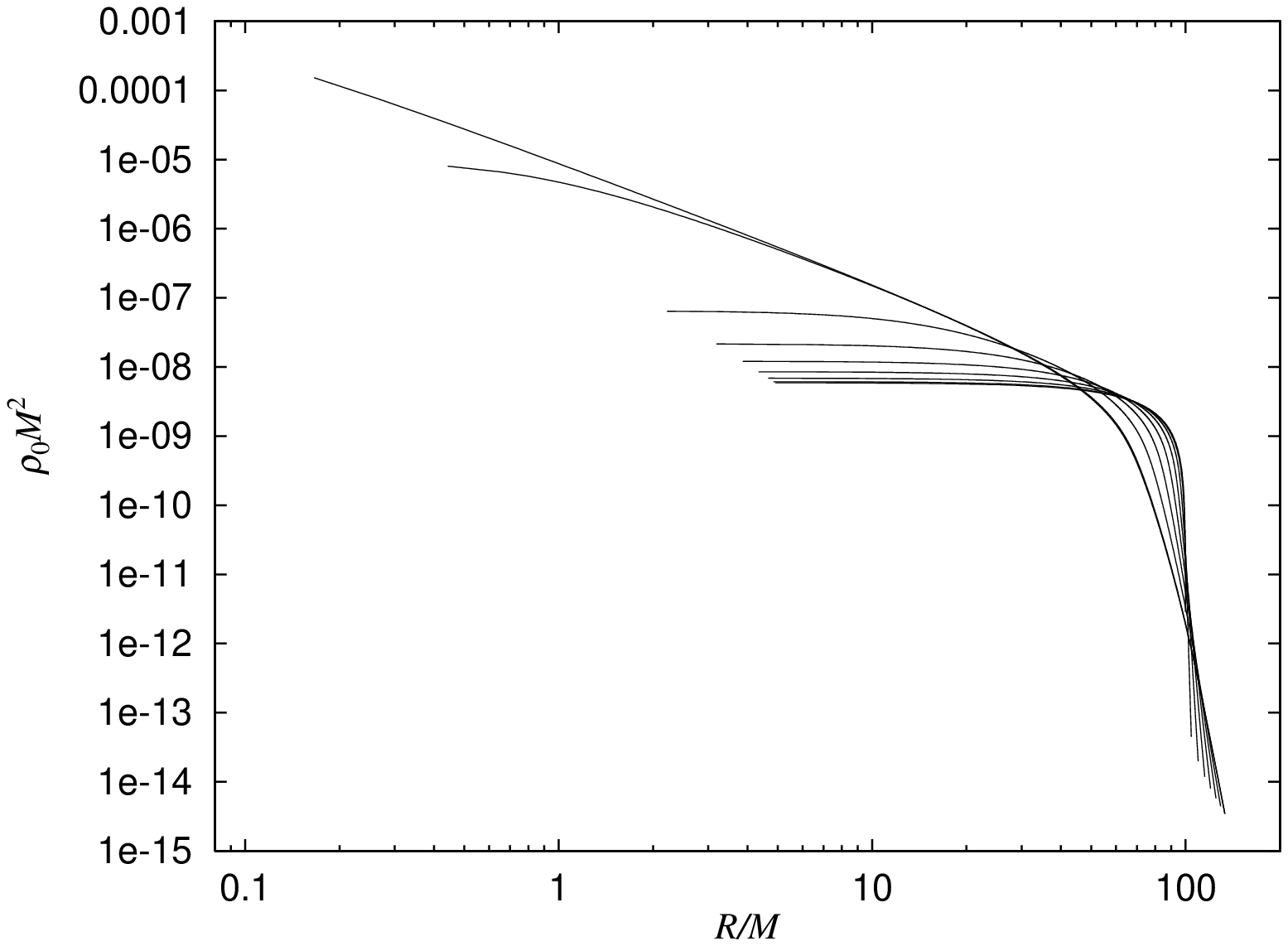}}
      \caption{}
      \label{fg:nsdems}
      \vspace{1cm}
      \centerline{\epsfysize 6cm \epsfxsize 9cm \epsfbox{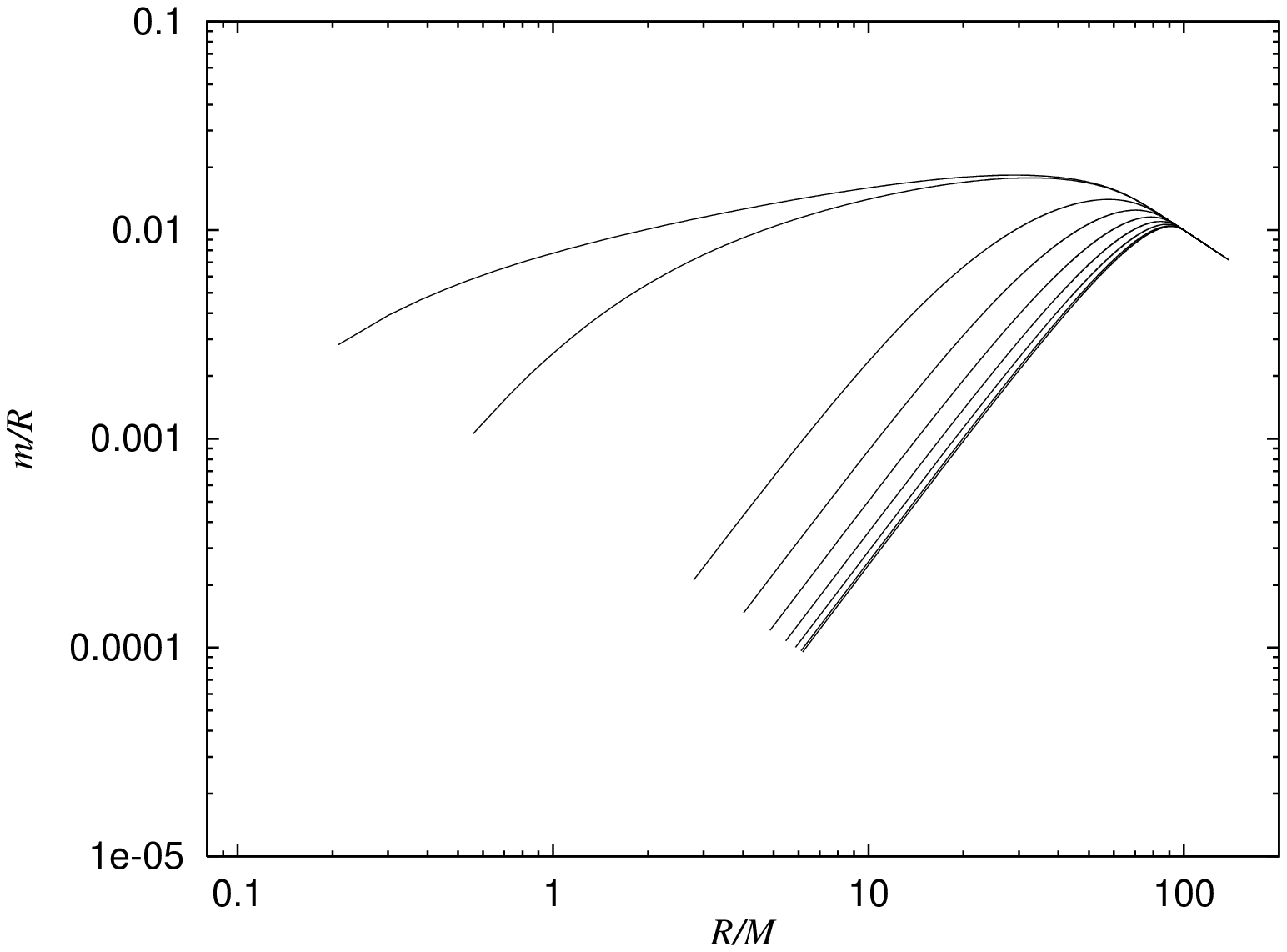}}
      \caption{}
      \label{fg:nsmrms}
      \vspace{1cm}
      \centerline{\epsfysize 6cm \epsfxsize 9cm \epsfbox{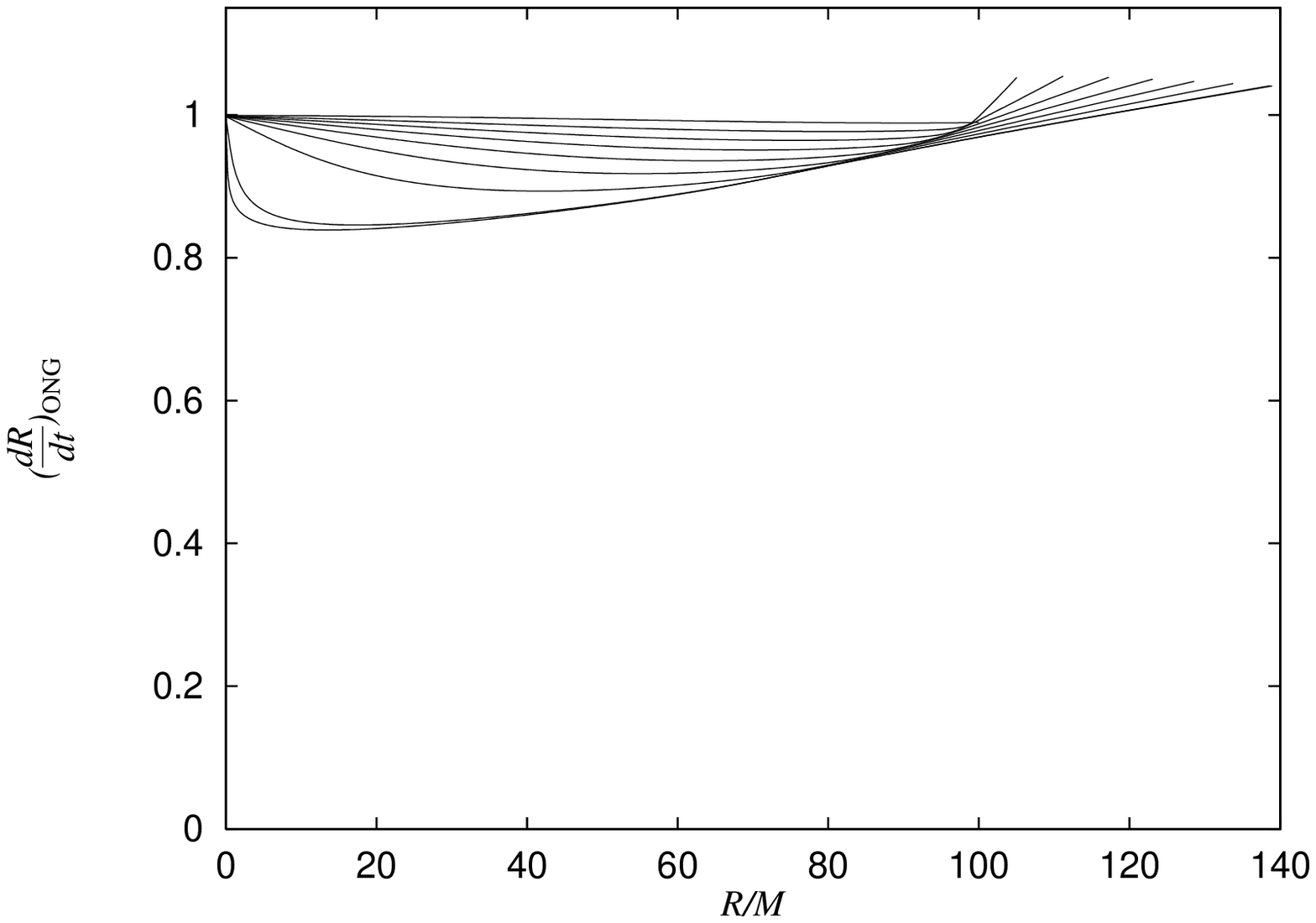}}
      \caption{}
      \label{fg:nsexms}
      \vspace{1cm}
      \centerline{\epsfysize 6cm \epsfxsize 9cm \epsfbox{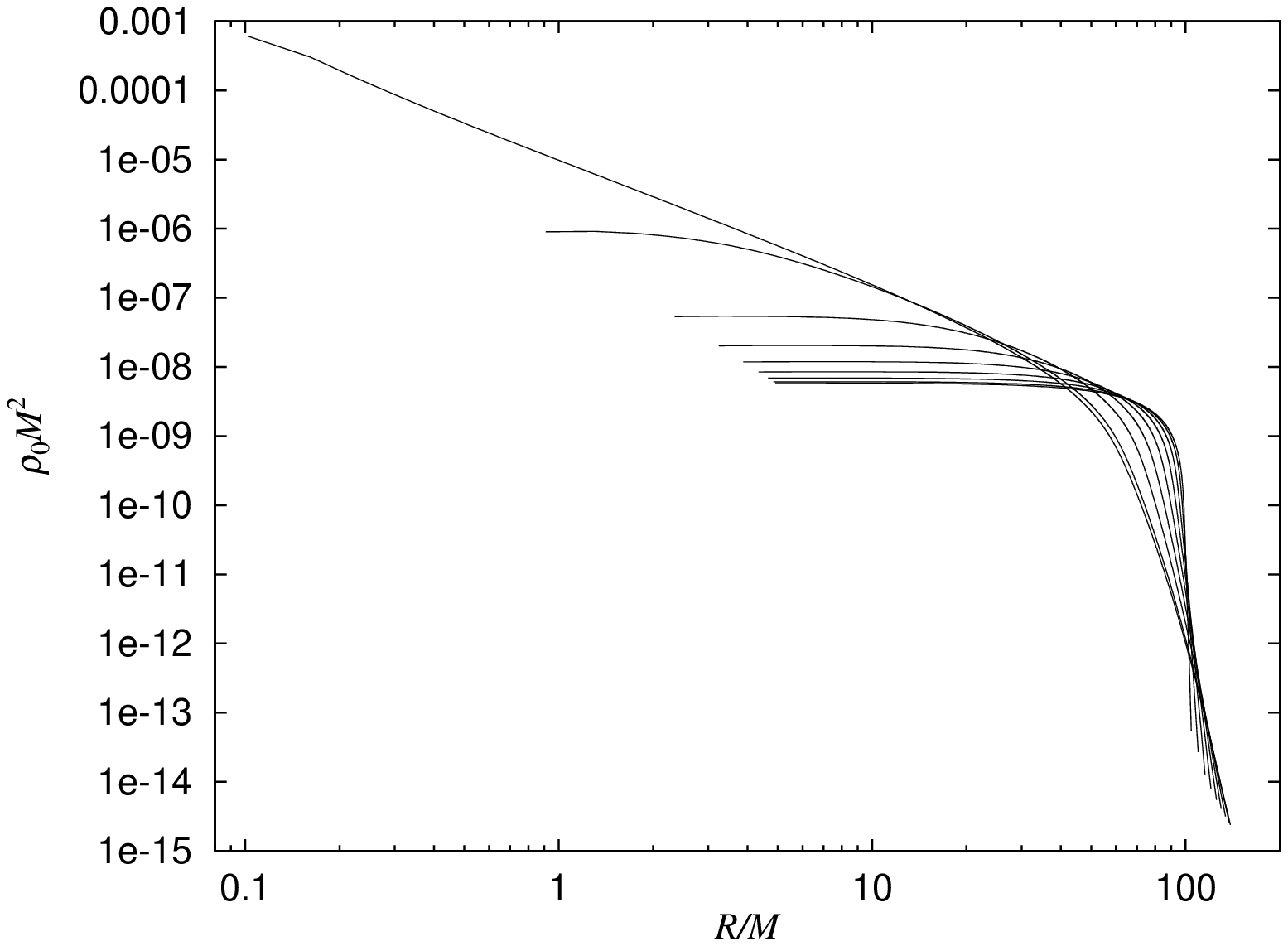}}
      \caption{}
      \label{fg:nsdehm}
      \vspace{1cm}
      \centerline{\epsfysize 6cm \epsfxsize 9cm \epsfbox{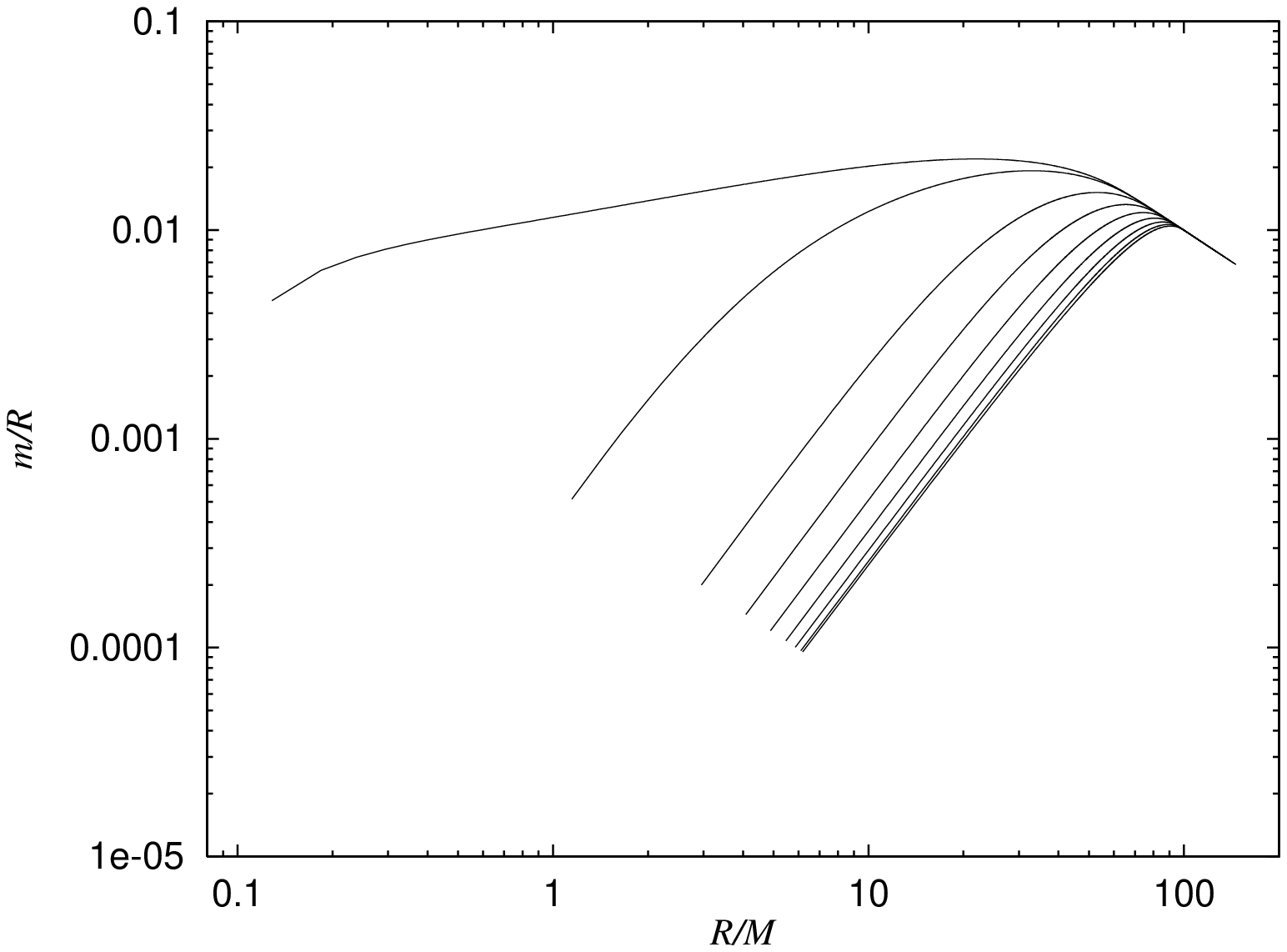}}
      \caption{}
      \label{fg:nsmrhm}
      \vspace{1cm}
      \centerline{\epsfysize 6cm \epsfxsize 9cm \epsfbox{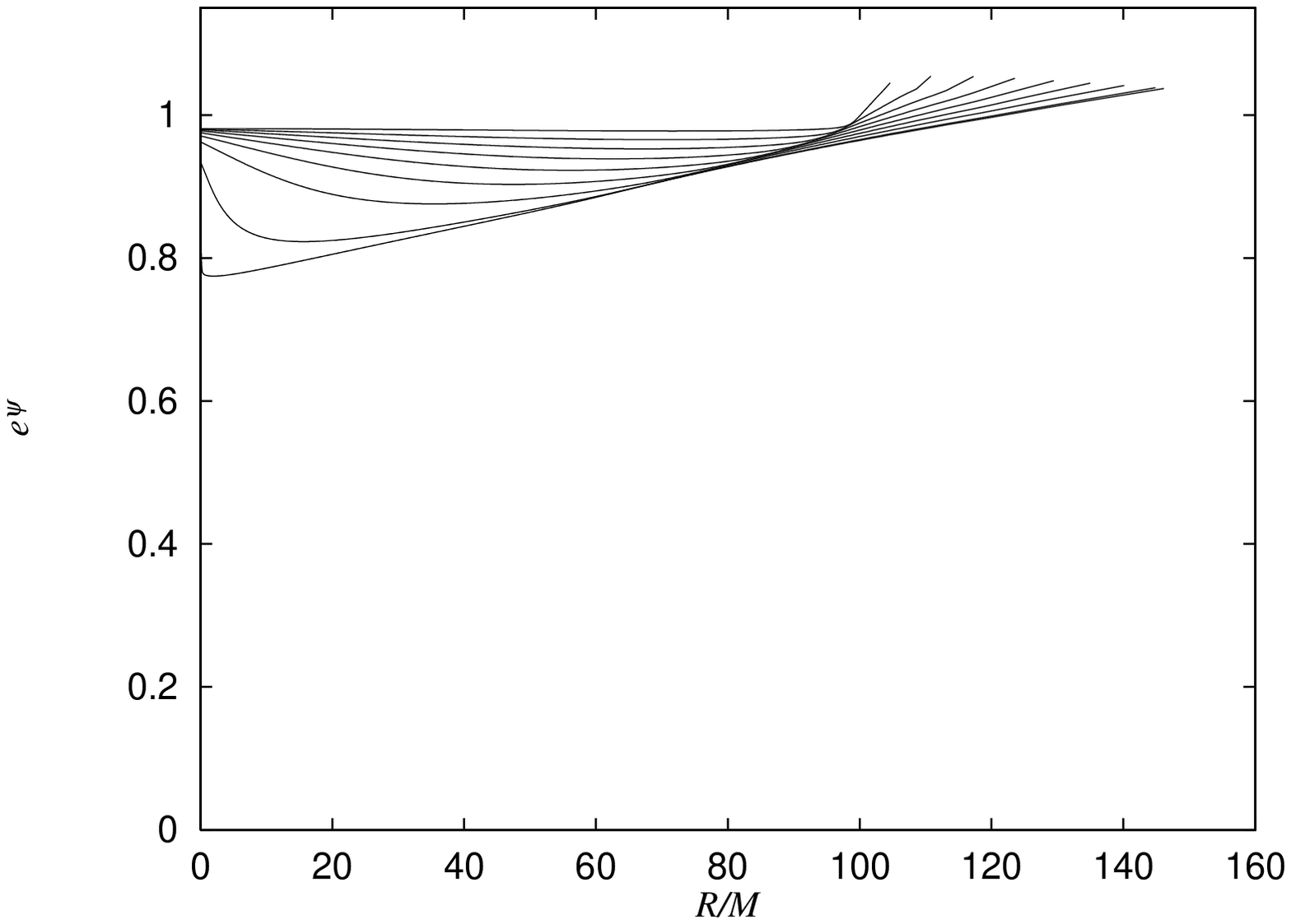}}
      \caption{}
      \label{fg:nslahm}
      \vspace{1cm}
      \centerline{\epsfysize 6cm \epsfxsize 9cm \epsfbox{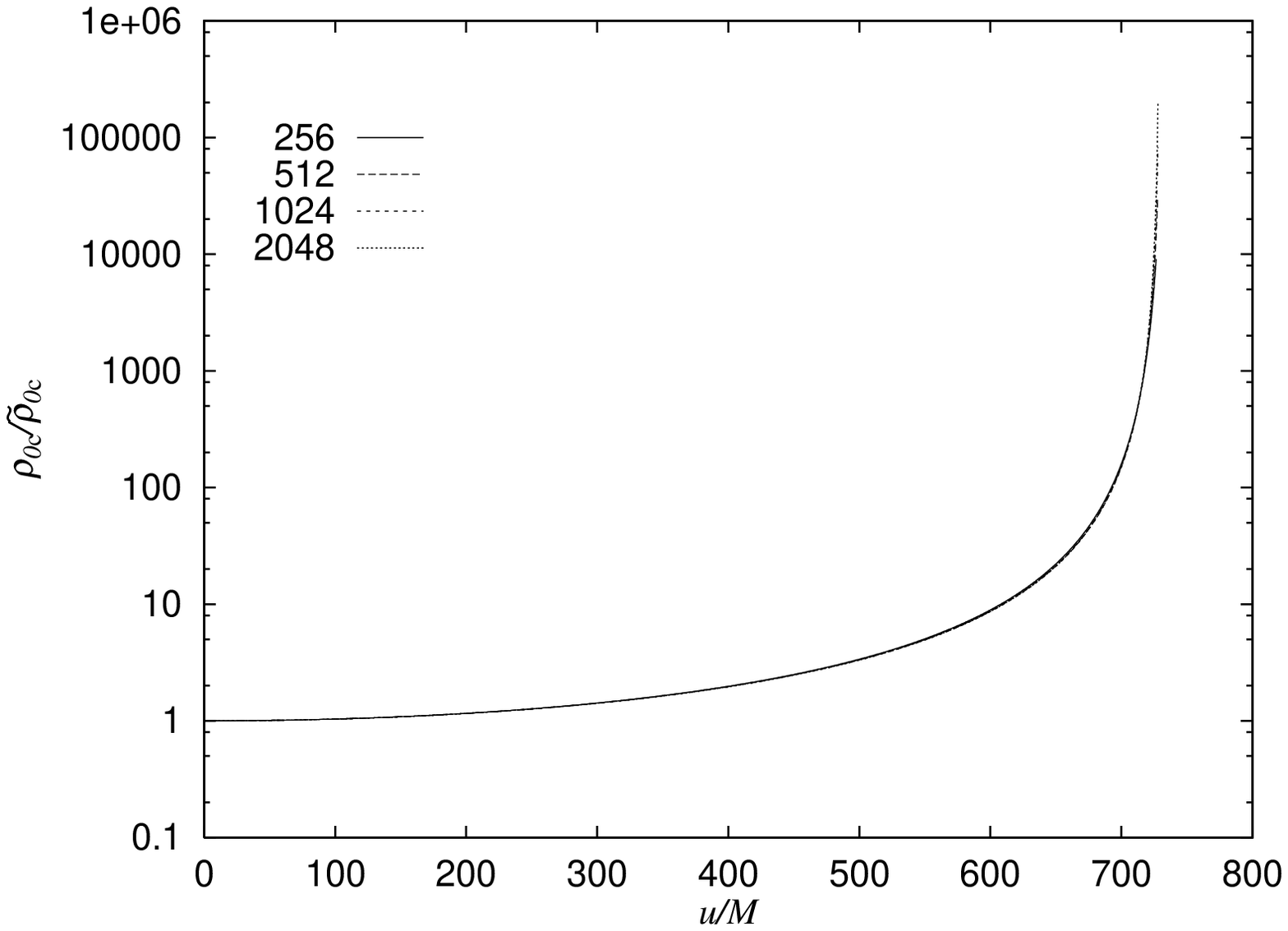}}
      \caption{}
      \label{fg:nscdhm}
      \vspace{1cm}
      \centerline{\epsfysize 6cm \epsfxsize 9cm \epsfbox{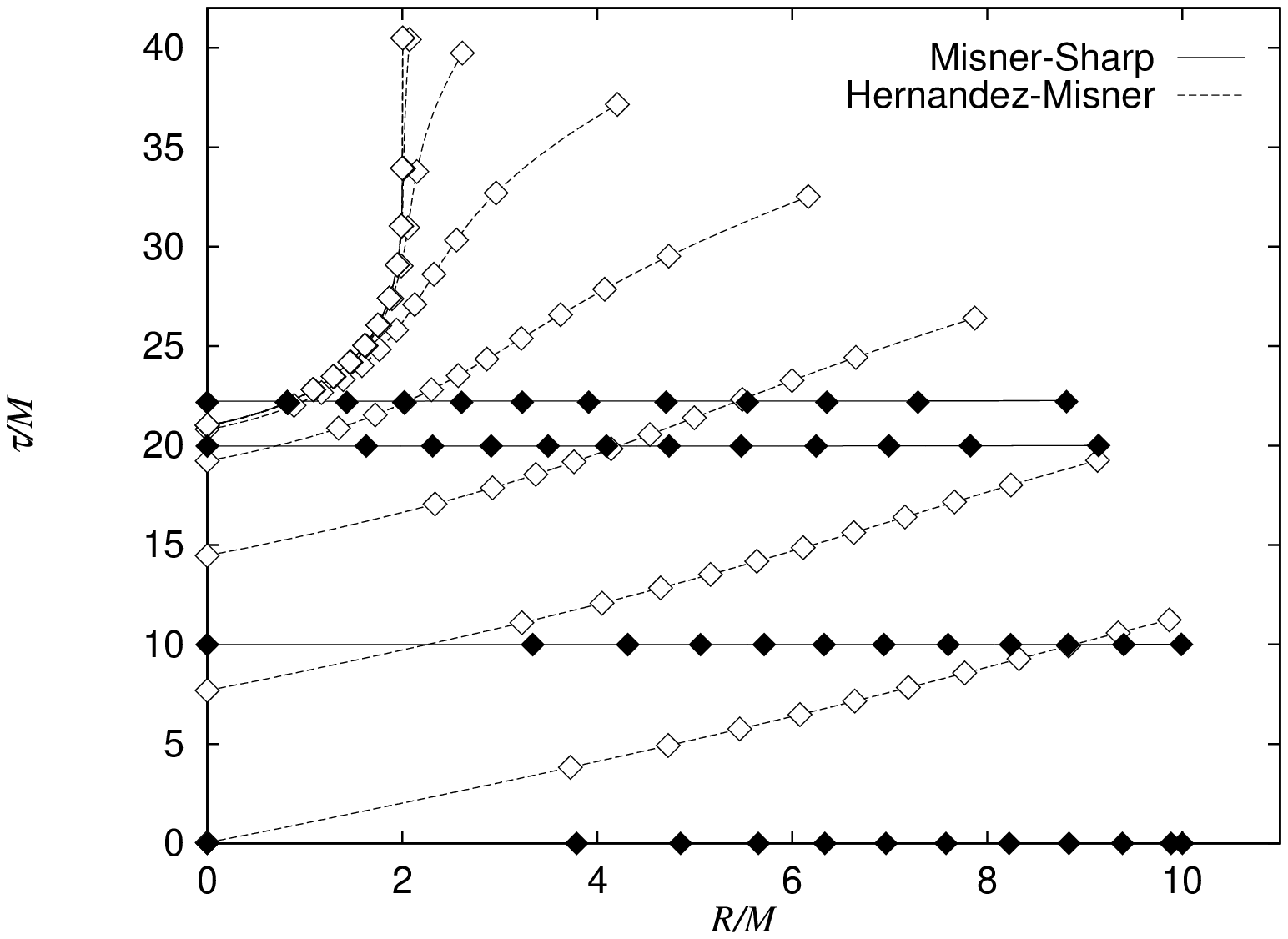}}
      \caption{}
      \label{fg:bhslmshm}
      \vspace{1cm}
      \centerline{\epsfysize 6cm \epsfxsize 9cm \epsfbox{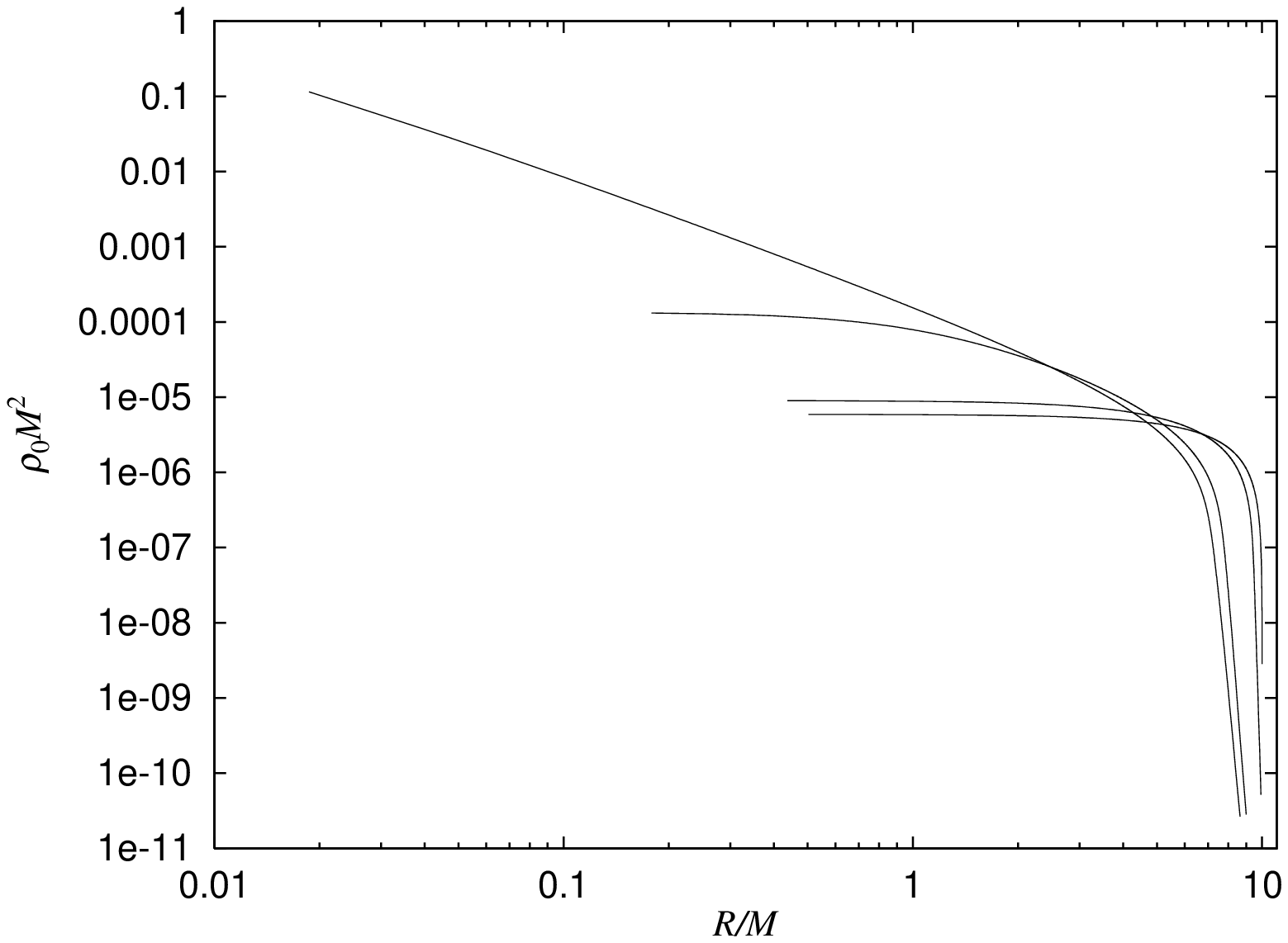}}
      \caption{}
      \label{fg:bhdems}
      \vspace{1cm}
      \centerline{\epsfysize 6cm \epsfxsize 9cm \epsfbox{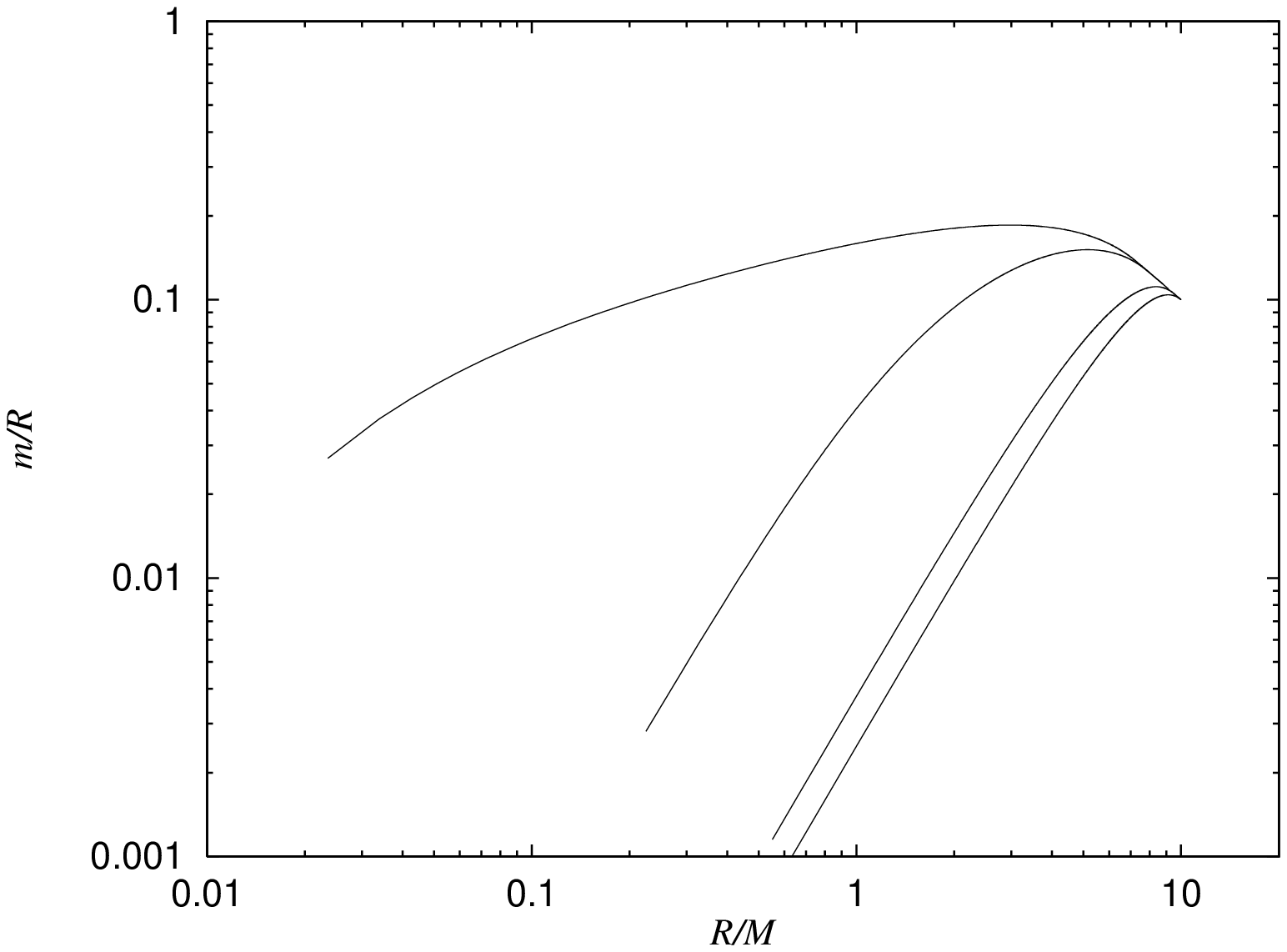}}
      \caption{}
      \label{fg:bhmrms}
      \vspace{1cm}
      \centerline{\epsfysize 6cm \epsfxsize 9cm \epsfbox{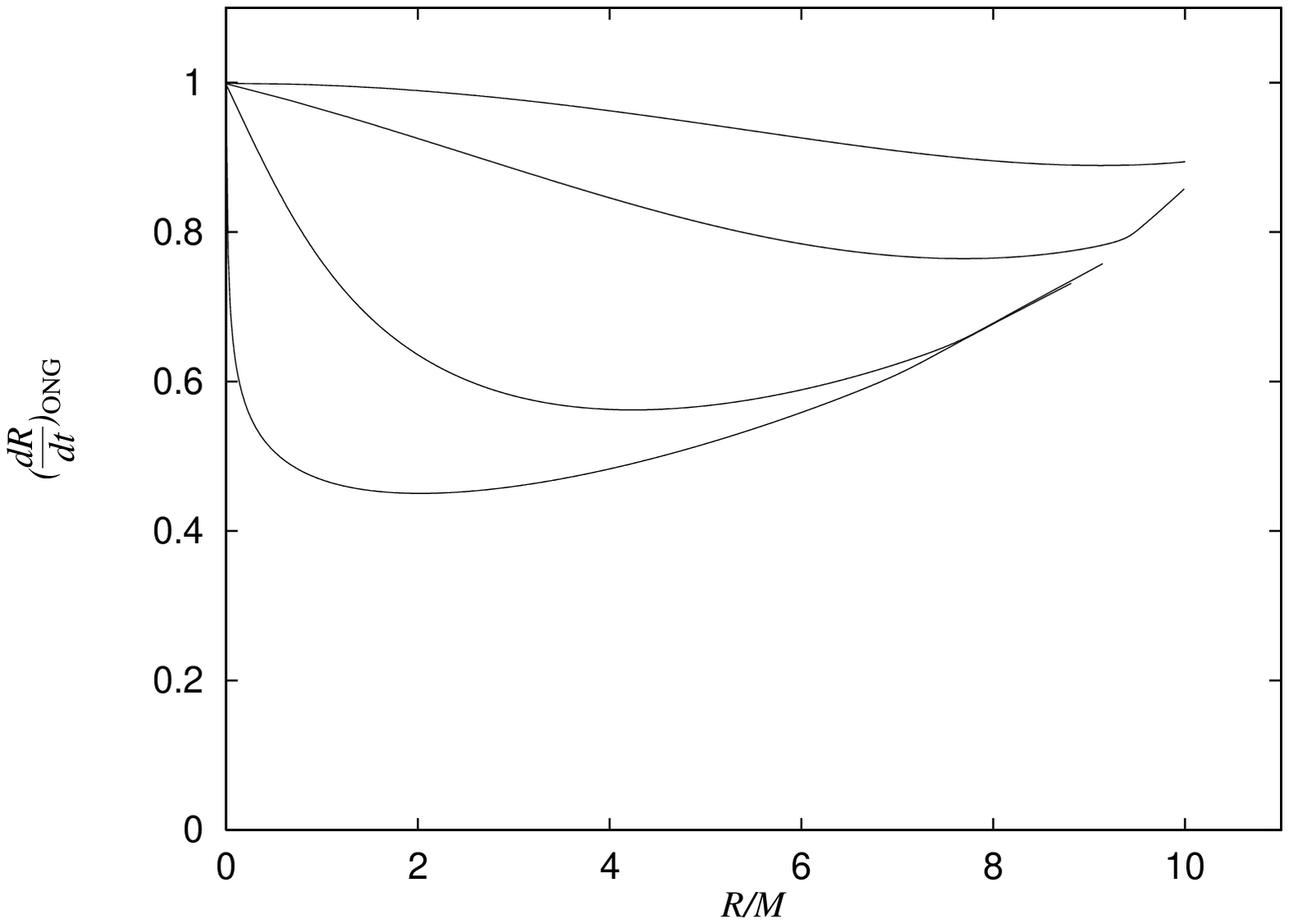}}
      \caption{}
      \label{fg:bhexms}
      \vspace{1cm}
      \centerline{\epsfysize 6cm \epsfxsize 9cm \epsfbox{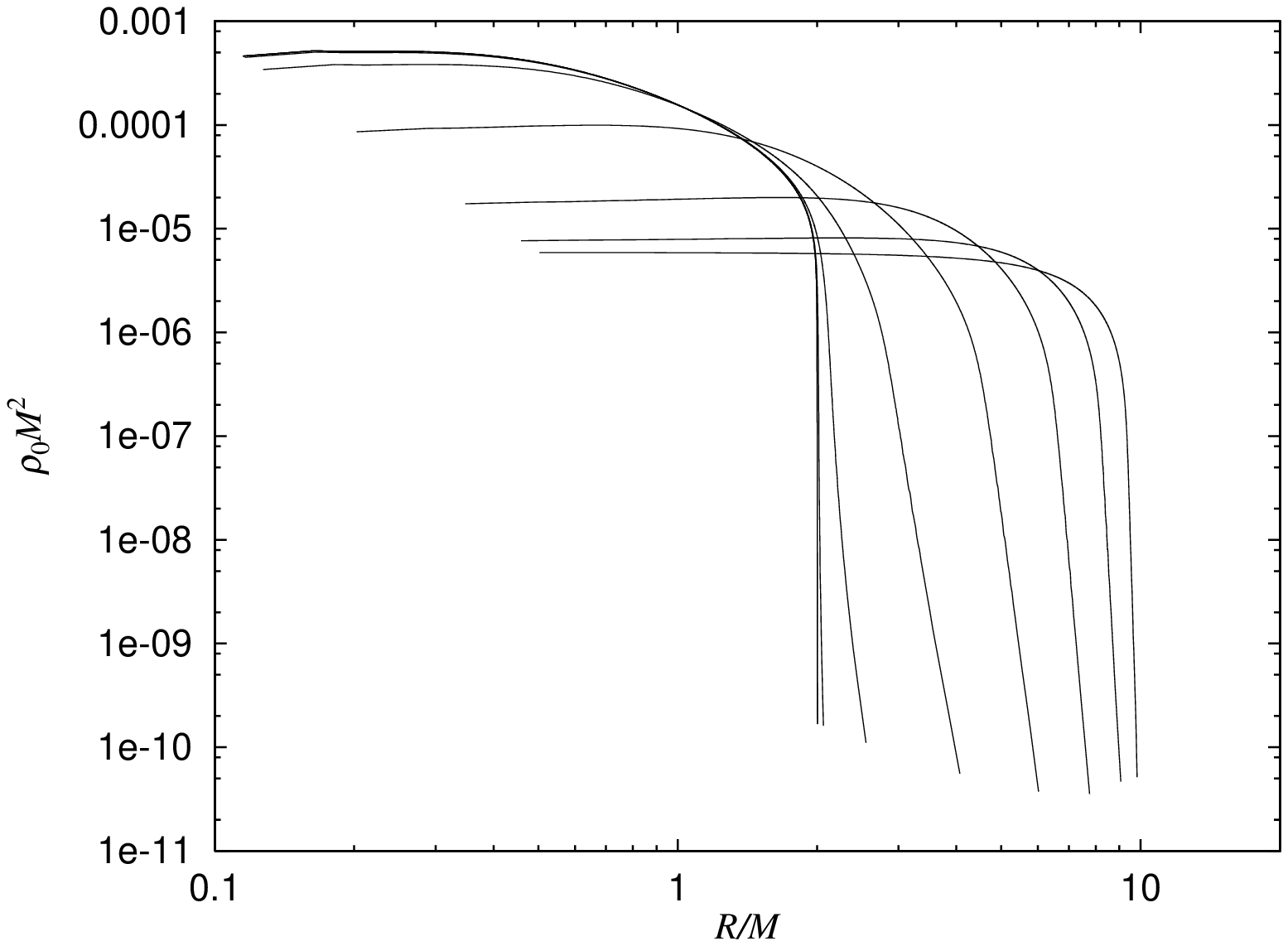}}
      \caption{}
      \label{fg:bhdehm}
      \vspace{1cm}
      \centerline{\epsfysize 6cm \epsfxsize 9cm \epsfbox{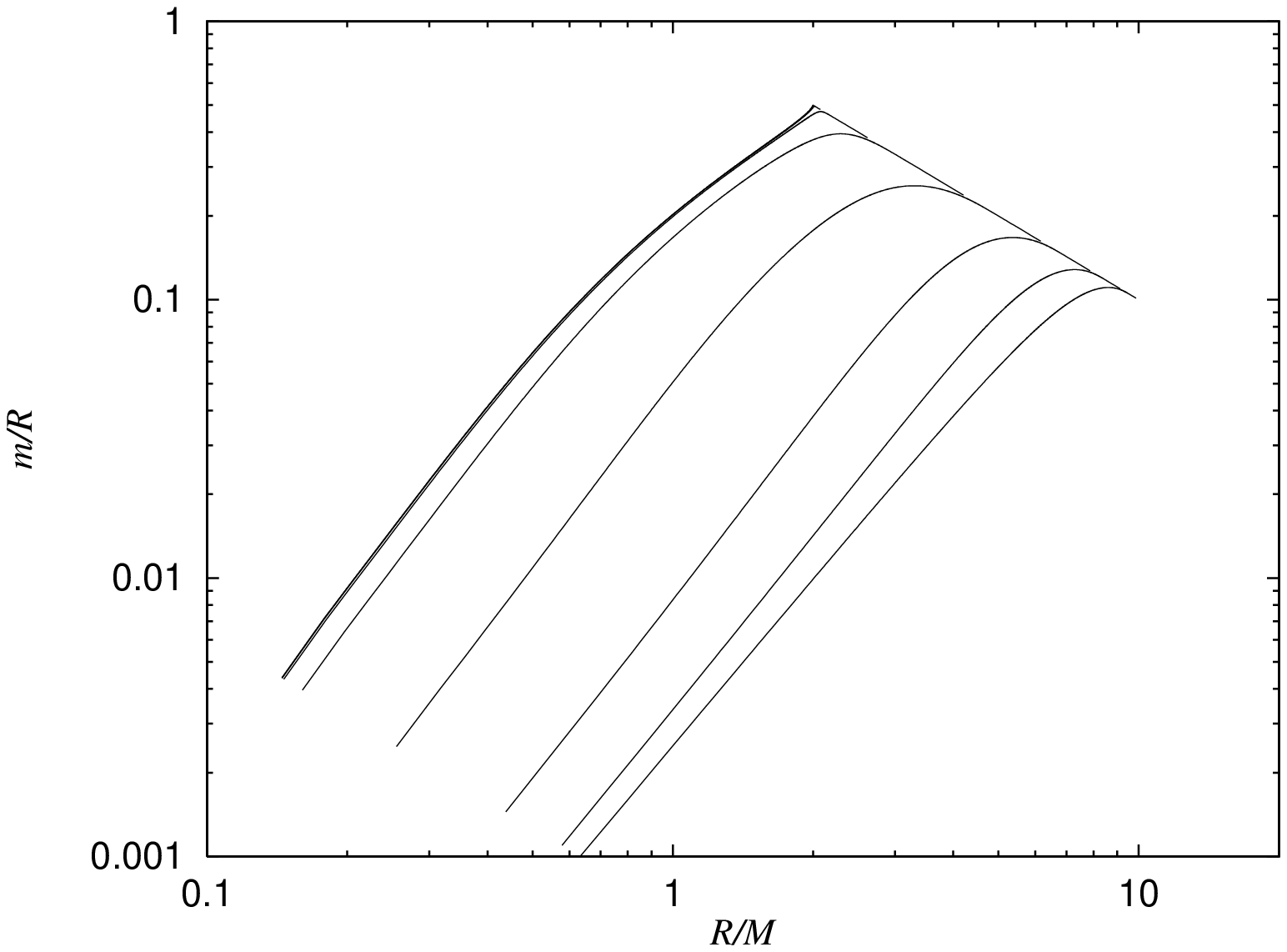}}
      \caption{}
      \label{fg:bhmrhm}
      \vspace{1cm}
      \centerline{\epsfysize 6cm \epsfxsize 9cm \epsfbox{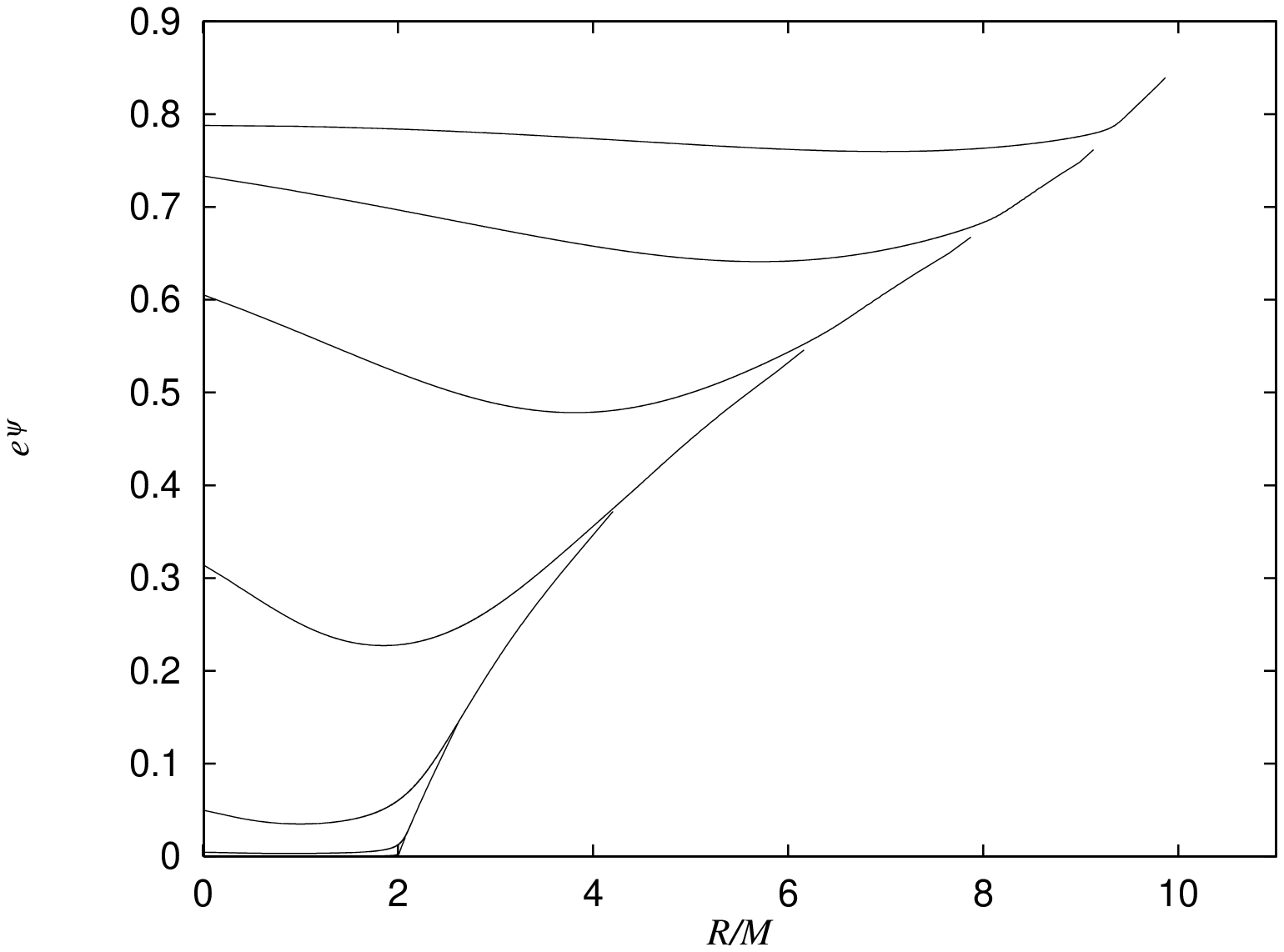}}
      \caption{}
      \label{fg:bhlahm}
      \vspace{1cm}
      \centerline{\epsfysize 6cm \epsfxsize 9cm \epsfbox{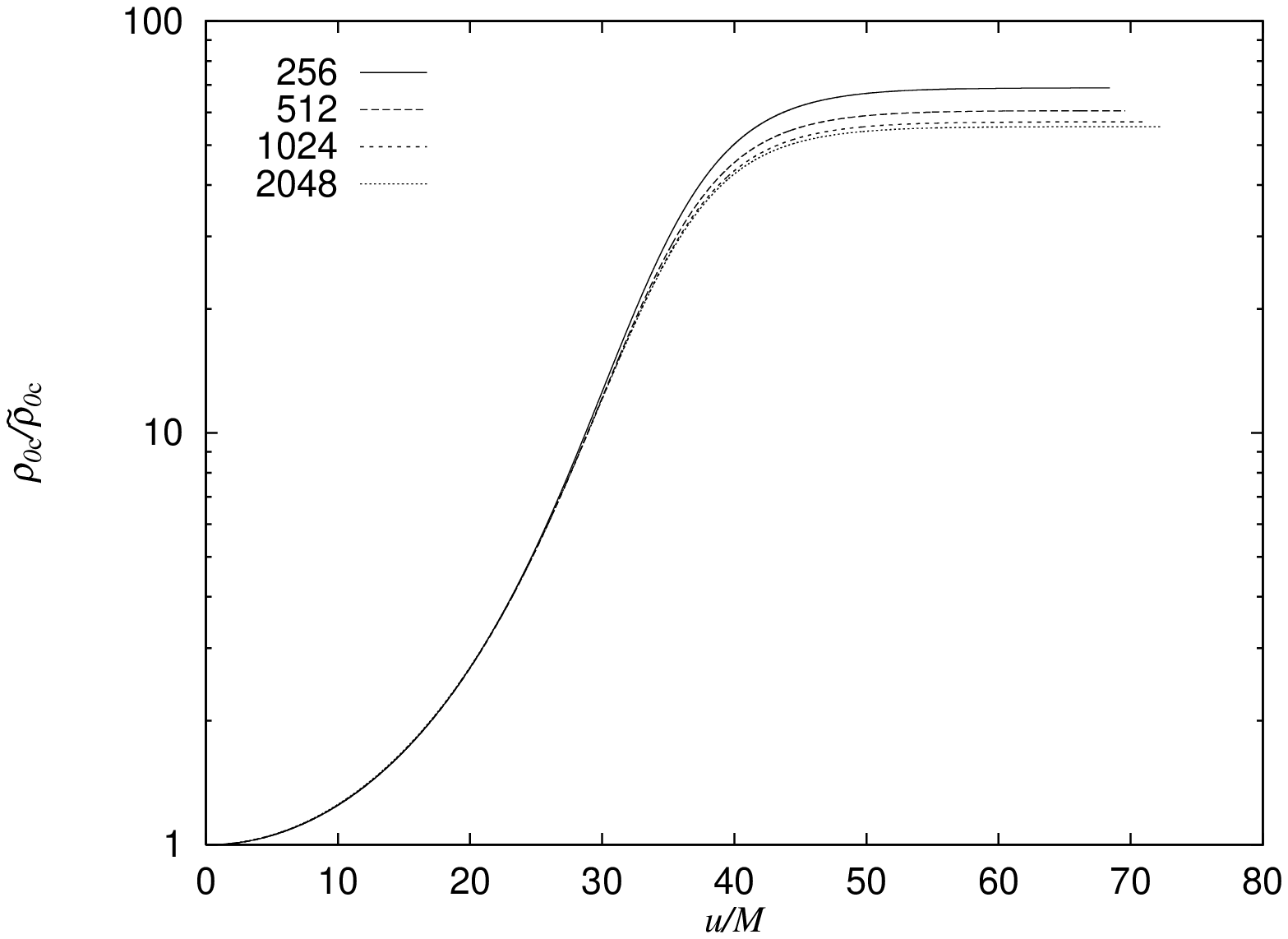}}
      \caption{}
      \label{fg:bhcdhm}
      \vspace{1cm}
      \centerline{\epsfysize 6cm \epsfxsize 9cm \epsfbox{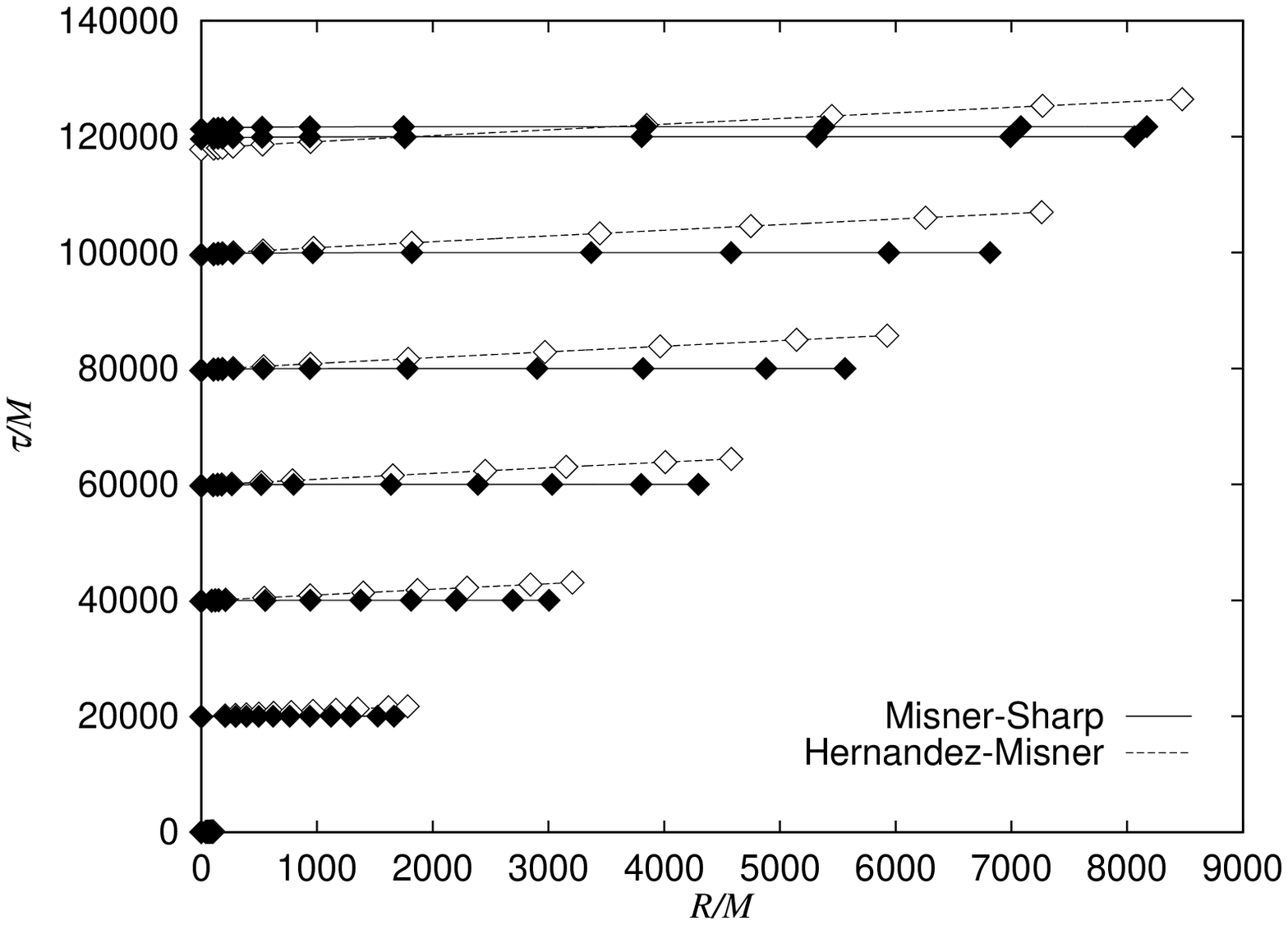}}
      \caption{}
      \label{fg:stslmshm}
      \vspace{1cm}
      \centerline{\epsfysize 6cm \epsfxsize 9cm \epsfbox{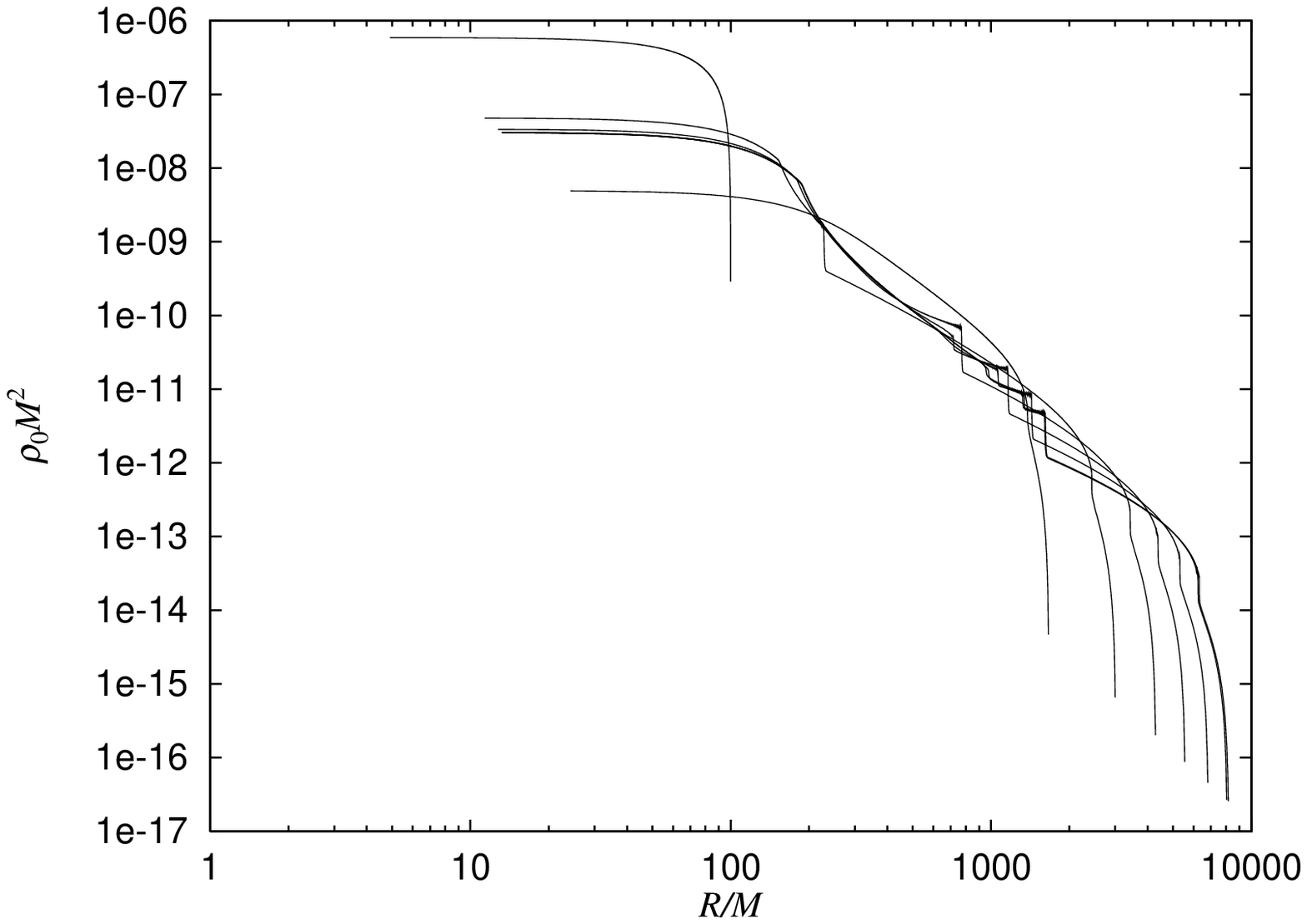}}
      \caption{}
      \label{fg:stdems}
      \vspace{1cm}
      \centerline{\epsfysize 6cm \epsfxsize 9cm \epsfbox{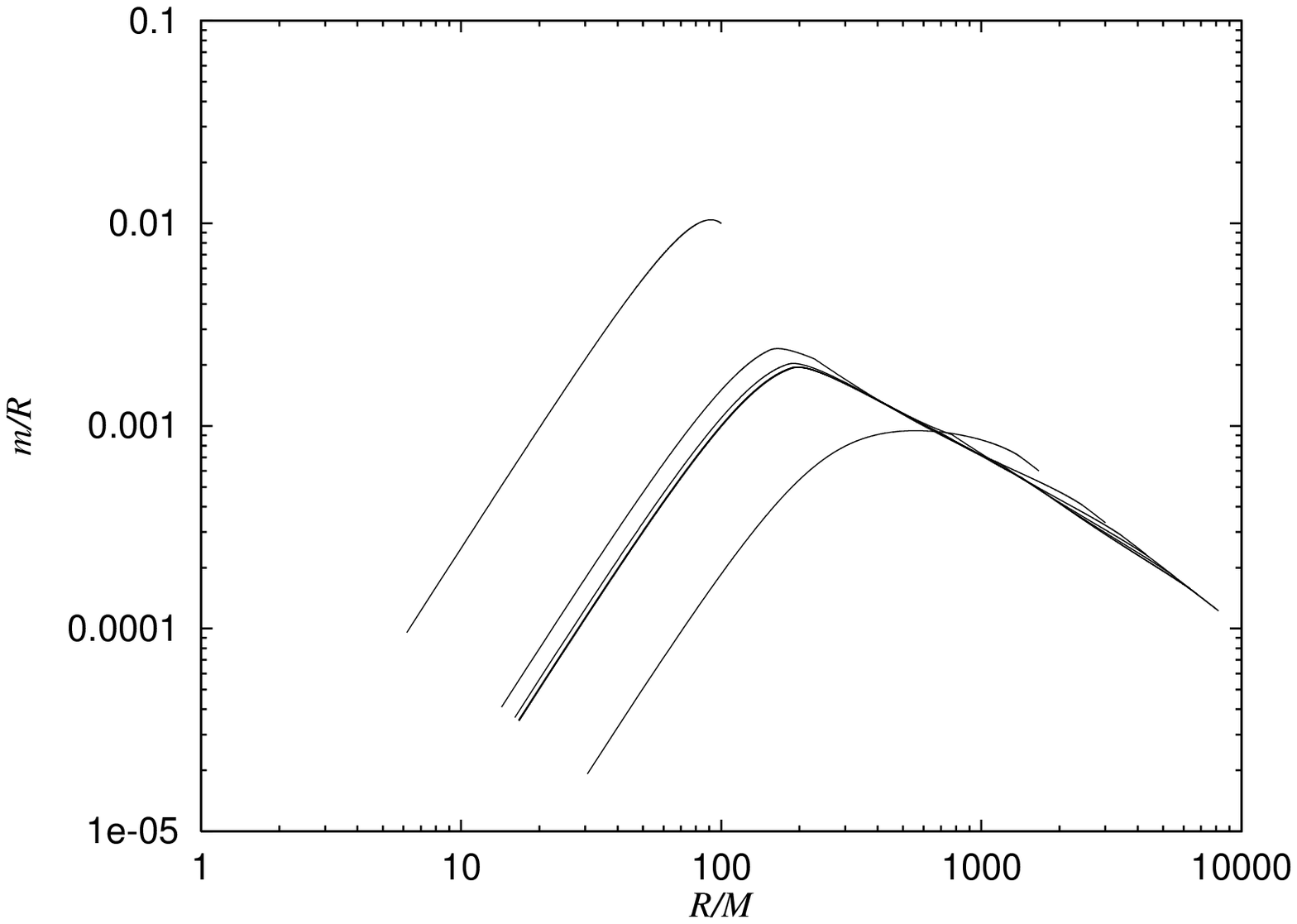}}
      \caption{}
      \label{fg:stmrms}
      \vspace{1cm}
      \centerline{\epsfysize 6cm \epsfxsize 9cm \epsfbox{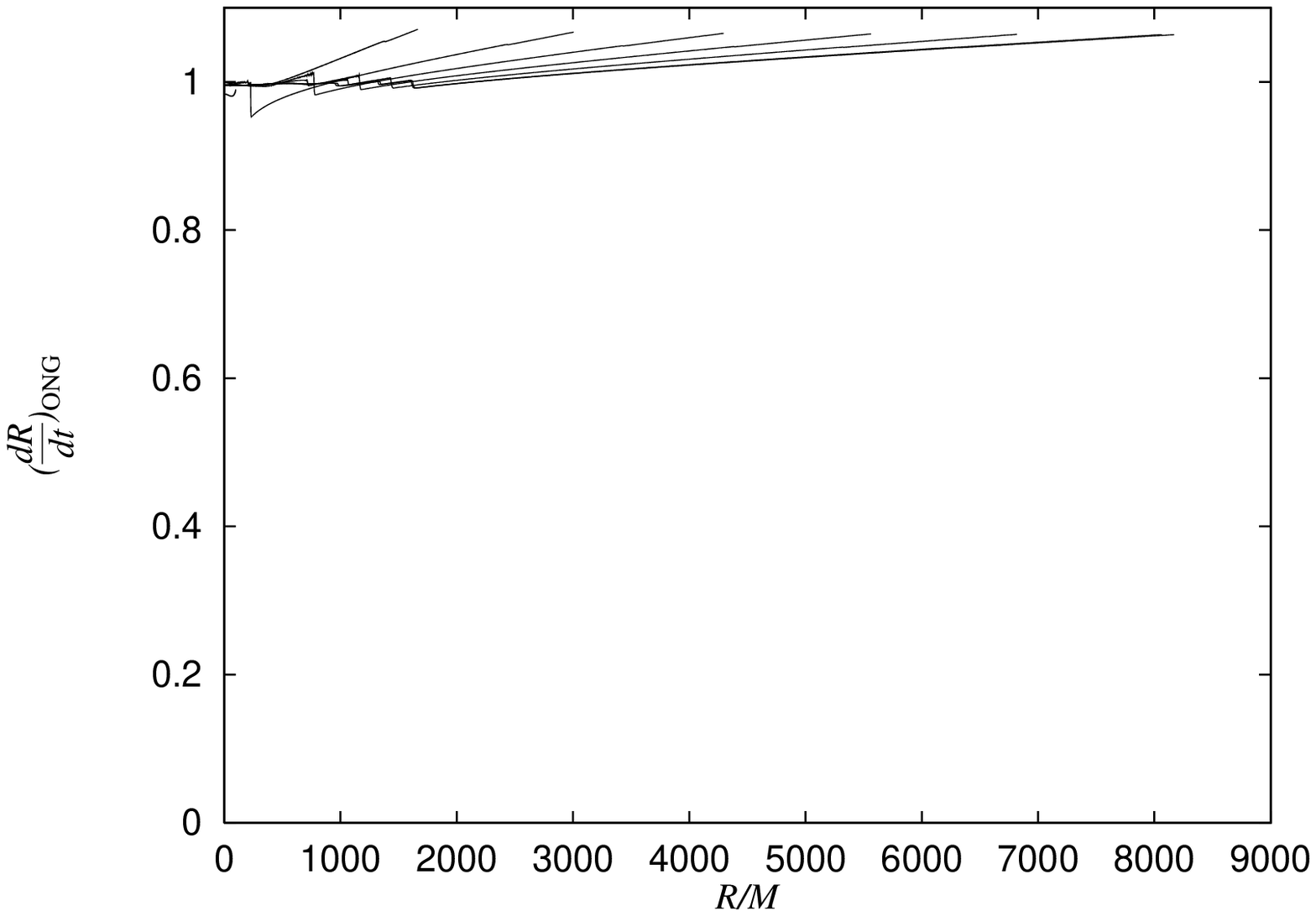}}
      \caption{}
      \label{fg:stexms}
      \vspace{1cm}
      \centerline{\epsfysize 6cm  \epsfxsize 9cm \epsfbox{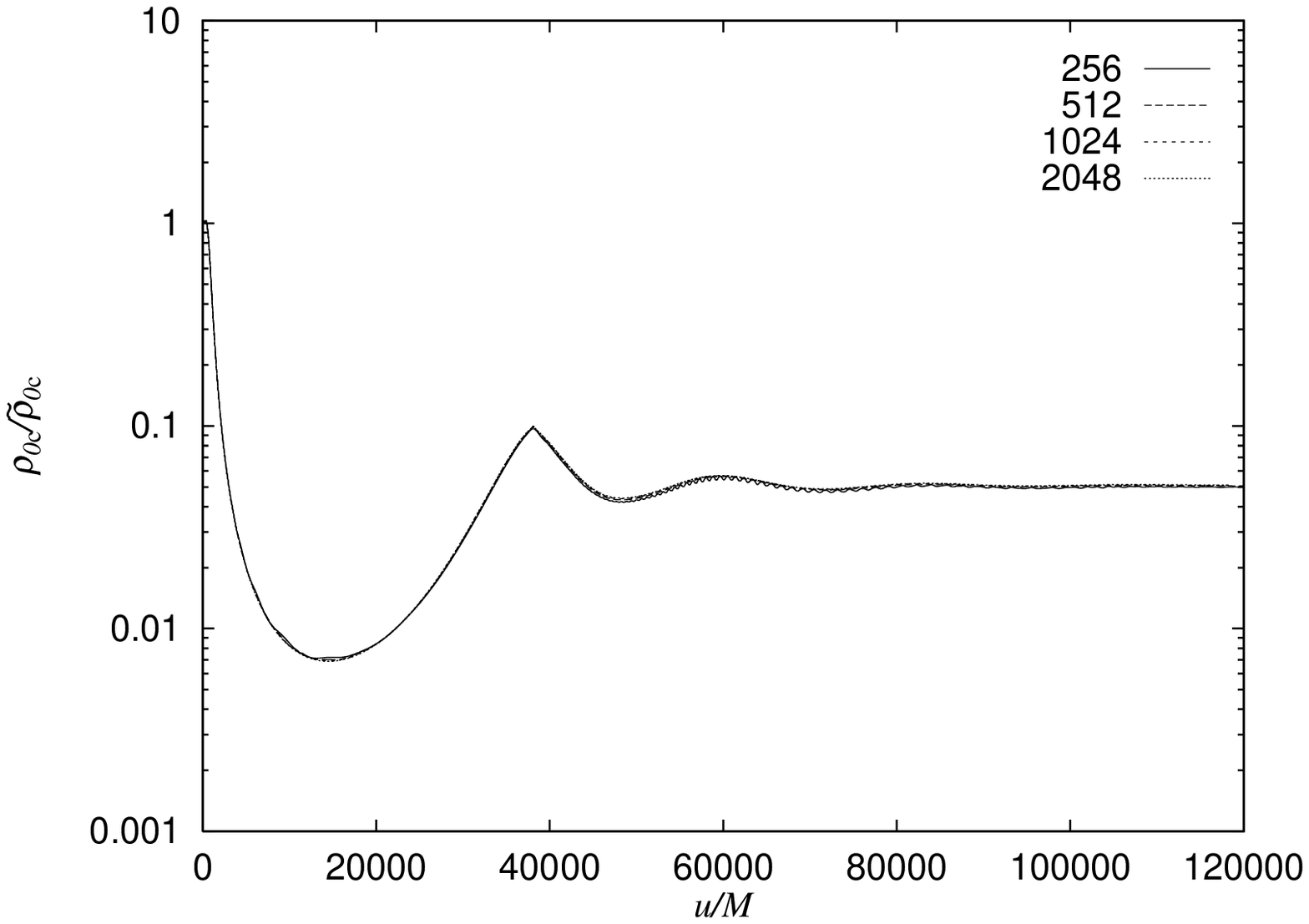}}
      \caption{}
      \label{fg:stcdhm}
\end{figure}

First we pay attention to the naked-singular case, the model in which
$\gamma-1=10^{-4}$, $\tilde{e}_{c}=10^2$, and $R_s=100M$,
where $M$ is the total gravitational mass.
Since the fluid is highly relativistic,
the equation of state is approximately
equivalent with the equation of state (\ref{eq:opeos}).
Hence, it is expected that the feature of collapse
is not sensitive to the value of $\tilde{e}_{c}$ if $\tilde{e}_{c}\gg 1$. 
In this calculation, the artificial viscosity was switched off.
Figure \ref{fg:nsslmshm} 
shows time slicing by the Misner-Sharp and
Hernandez-Misner codes.
The ordinate is the proper time $\tau$ of a comoving observer,
and the abscissa is the circumferential radius. 
The Misner-Sharp slicing presented in Fig. \ref{fg:nsslmshm}
is a family of spacelike hypersurfaces
$t/M=0, 100, 200, 300, 400, 500, 600, 700, 707$,
where the rescaling freedom of $t$ is fixed so that
$t$ agrees with the proper time at the stellar surface.
On the last slice $t=707M$, the Misner-Sharp code 
detected a central singularity based on the criteria
described above.
The Hernandez-Misner slicing is a family of null hypersurfaces
$u/M=0,100, 200, 300, 400, 500, 600, 700, 728$.
Also on the last slice $u=728M$, a central singularity was detected.
In this figure, locations of some fluid elements are
also marked.  
Figures \ref{fg:nsdems}--\ref{fg:nsexms} 
shows the Misner-Sharp 
time evolution of the rest-mass density $\rho_0$, 
the ratio $m/R$, 
and $dR/dt$ along outgoing null geodesics
[which is hereafter denoted as $(dR/dt)_{ONG}$],
where $m$ is the Misner-Sharp mass~\cite{ms1964}.
As seen in Fig. \ref{fg:nsdems}, 
the time evolution of the density profile in this model
looks similar to the Penston's dust collapse 
solution in Newtonian gravity~\cite{penston1969}
and also the LTB solution in Einstein gravity.
It is remarkable that the density distribution in the central region
approaches a power-law profile 
and therefore loses any characteristic scale.
From Fig. \ref{fg:nsdems}, the divergent 
behavior of the density at the center 
with respect to $R$ changes at the occurrence 
of the singularity as
\begin{equation}
  \rho_0\propto \mbox{const} \Longrightarrow \rho_0\propto R^{-\alpha},
\end{equation}
where $\alpha\simeq 1.7$.
Penston~\cite{penston1969} showed that $\alpha=12/7$ for the dust
collapse in Newtonian gravity.
In the Appendix, we will show that 
$\alpha=12/7$ is also valid 
for the LTB solution on the spacelike hypersurface
$t=\mbox{const}$ of the occurrence of the central singularity.
As seen in Fig. \ref{fg:nsmrms}, 
the ratio $m/R$ is much less than unity.
This suggests that this collapse is well approximated 
by that in Newton gravity.
The behavior of the ratio changes 
at the occurrence of the singularity
\begin{equation}
  \frac{m}{R}\propto R^2 \Longrightarrow \frac{m}{R}\propto R^{\beta},
\end{equation}
where $\beta\simeq 0.3$.
It should be also noted that $\beta=2/7$ in
the Penston's dust collapse 
solution. 
In the Appendix, we will show that 
$\beta=2/7$ for the LTB solution.
Figure \ref{fg:nsexms} shows that the expansion of outgoing 
null geodesics are always positive
until the central singularity is detected.
In other words, the Misner-Sharp code does not find the
apparent horizon before the occurrence of the singularity. 
Figures \ref{fg:nsdehm}--\ref{fg:nslahm} shows Hernandez-Misner 
time evolution of $\rho_0$, $m/R$, 
and the lapse function $e^{\psi}$.
From Figs. \ref{fg:nsdems} and \ref{fg:nsdehm}, it is found that
there is little difference 
in the divergence property of the density profile 
in the central region in both codes. 
In the Appendix, we will show that $\alpha=12/7$ and $\beta=2/7$ 
for the LTB solution 
also on the earliest null ray 
which emanates from the central naked singularity. 
Figure \ref{fg:nsmrhm} 
shows that the ratio $m/R$ is much less than unity
also on that null ray.
In Fig. \ref{fg:nslahm},
it is found that 
$e^{\psi}$ does not vanish but remains of the order
of unity until the central singularity is detected.
Since $e^{\psi}$ remains of the order of unity,
an event horizon has not yet formed.
Figure \ref{fg:nscdhm} shows the growth of the central rest-mass density 
$\rho_{0c}$ in this model.
The simulation was repeated with various radial grid resolutions.
Each curve is labeled by the number of spatial grid zones used.
The value of the central rest-mass density grows
unboundedly.
The blow up of the central rest-mass density 
becomes more rapid and the maximum value of it
that can be attained becomes larger as 
the resolution becomes higher.
Then, in summary, the collapse is well approximated by 
dust collapse both in Newton gravity and in Einstein gravity,
and a central naked singularity
forms in this model based on the present criteria.

\subsection{black hole}
Next we take the model in which 
$\gamma-1=10^{-4}$, $\tilde{e}_{c}=10^2$, and $R_s=10M$
as an example of black hole formation.
Also in this calculation, the artificial viscosity was
switched off.
Figure \ref{fg:bhslmshm} shows time slicing by the Misner-Sharp and
the Hernandez-Misner codes.
The former slicing is $t/M=0,10,20,22.3$
and the latter is $u/M=0,10,20,30,40,50,60,70,72.3$.
The former code was stopped because of the steepness
of the density profile around the center,
while the latter code was stopped because 
$e^{\psi}$ became less than $10^{-3}$
all over the star.
The sequence of the outgoing null geodesics $u=\mbox{const}$ 
converges, and its limit curve is an event horizon.
Figures \ref{fg:bhdems}--\ref{fg:bhexms}
shows the Misner-Sharp time evolution of $\rho_0$, $m/R$, 
and $(dR/dt)_{ONG}$.
The behavior of $\rho_0$ and $m/R$ in the central region seen 
in Figs. \ref{fg:bhdems} and \ref{fg:bhmrms} 
is quite similar to that seen in the naked-singular case.
From Fig. \ref{fg:bhmrms}, the ratio $m/R$ is not so small although
it remains less than $1/2$ which corresponds to the apparent horizon
in the spherically symmetric space-time.
Figure \ref{fg:bhexms} shows that the Misner-Sharp code 
does not detect the apparent horizon until the occurrence
of the central singularity
although the singularity is covered by 
the event horizon.
Figures \ref{fg:bhdehm}--\ref{fg:bhlahm} 
shows the Hernandez-Misner time evolution of $\rho_0$,
$m/R$, and $e^{\psi}$.
It is seen in Fig. \ref{fg:bhdehm} 
that the density profile around the center 
is not so steep even at the event horizon.
The ratio $m/R$ is increased and reaches $1/2$ at the surface.
Therefore the Newtonian approximation is not valid.
In Fig. \ref{fg:bhlahm}, it is shown that $e^{\psi}$ approaches zero
as $u$ increases, which indicates approach to the event horizon. 
Figure \ref{fg:bhcdhm} shows growth of 
the rest-mass density at the center for this case.
The simulation was repeated with various radial grid resolutions.
From this figure it would be sure that the resolution is sufficient
for the following conclusion.
Since the Hernandez-Misner code detects an event horizon
before the occurrence of a central singularity,
the singularity is covered by the event horizon.
Moreover, it can be expected 
that this collapse would result in the locally naked
singularity because the Misner-Sharp time evolution in the central
region is very similar to that of the globally naked singular case.

\subsection{stable star}
The final fate of collapse for $\gamma=5/3$, 
$\tilde{e}_c=10^{-2}$, and
$R_s=100M$ is a stable star.
In this calculation, the artificial viscosity was switched on
in order to suppress numerical instabilities around the shock front.
Figure \ref{fg:stslmshm} shows time slicing by both codes.
The Misner-Sharp slicing is $t/M=0,2.00\times 10^4$, $4.00 \times 10^4$,
$6.00 \times 10^4$, $8.00 \times 10^4$, $1.00\times 10^5$, $1.20\times 10^5$, 
$1.22 \times 10^5$.
The Hernandez-Misner slicing is $u/M=0$, 
$2.00\times 10^4$, $4.00 \times 10^4$,
$6.00 \times 10^4$, $8.00 \times 10^4$, 
$1.00 \times 10^5$, $1.18\times 10^5$.
In this figure, it is seen that the motion of the fluid is 
much slower than the speed of light.
Hence, the time evolutions both in the Misner-Sharp and
Hernandez-Misner time slicings are basically the same.
From this reason, we present here only the Misner-Sharp 
time evolution.
Figures \ref{fg:stdems}--\ref{fg:stexms} 
show the time evolution of $\rho_0$, $m/R$, and 
$(dR/dt)_{ONG}$ respectively.
In Fig. \ref{fg:stdems}, it is found that 
the stable star has the core-envelope structure.
The surface of the envelope keeps expanding, while
the core is accreting the envelope.
The core does not collapse but settles its density profile 
after many dynamical time scales.
As seen in Figs. \ref{fg:stdems} and \ref{fg:stmrms},
the resultant core radius is about $200M$,
consistent with initial total internal energy.
Figure \ref{fg:stmrms} shows that the Newtonian approximation
is valid because the ratio $m/R$ is much less than unity.
The fact that $m/R$ is proportional to $R^{-1}$
in the envelope
indicates that the mass contained in the envelope 
is negligibly small
compared to the core mass.
Of course, the expansion of the outgoing null geodesics
is always positive, as seen in Fig. \ref{fg:stexms}.
Figure \ref{fg:stcdhm} shows that the central region 
settles down after several oscillations.
The simulation was repeated with various radial grid resolutions.

\subsection{parameter search}
Tables \ref{tb:hr}--\ref{tb:nr} summarize the final fate of collapse
for $\tilde{e}_{c}=10^2,1,10^{-2}$.
Table \ref{tb:hrde} is the detailed search of the 
critical parameter region of Table \ref{tb:hr}.
B, N, E, BE, and SE, 
mean a black hole, a naked singularity,
an expansion, a black hole with an envelope, and 
a star with an envelope, respectively.
X and Y indicate some technical difficulties.
X means that the present method does not work since 
a central singularity occurs before the first ray
from the center reaches the stellar surface in the Misner-Sharp code.
Y means that the stellar surface goes outward so rapidly
that some numerical difficulty occurs.

From Tables \ref{tb:hr} and \ref{tb:hrde}, 
we find that the final fate of collapse
from less compact ($R_s/M\agt 20$) density distribution is, 
in general, not a black hole but a naked singularity
if the fluid is highly relativistic 
and $\gamma \alt 1.01$.
The final fate of collapse from compact ($R_s/M\alt 10$) 
density distribution
is a black hole even if the fluid is highly relativistic 
and $\gamma \alt 1.01$.
If the fluid is highly relativistic 
and $\gamma \agt 1.01$, the fluid begins to expand
from less compact density distribution.
It was confirmed that the above statements do not
depend on details of initial density profile.
From Table \ref{tb:nr},
we find that,
if the fluid is not relativistic and 
its profile is less compact,
the usual picture of collapse in Newton gravity is true.
If $\gamma<4/3$, the pressure gradient cannot sustain 
the gravitational collapse.
If $\gamma>4/3$, the final fate of collapse is 
a black hole, a naked singularity or a stable star
depending on
energetics, i.e., the total internal energy and the 
gravitational energy. 
It should be understood that, in Table \ref{tb:nr}, 
an N, i.e., a naked 
singularity means not $e\gg 1$ but only
the steep density profile at the center.
In fact, in the present calculation,
$e$ did not become much larger than the initial value
until the central singularity breaks down
our numerical code because of the finite resolution.
This suggests that the dynamical range of the code used here
is not sufficient to recognize an N in Table \ref{tb:nr} 
as a genuine naked singularity.
From Tables \ref{tb:hr}--\ref{tb:nr}, based on the criteria described above,
a naked singularity can occur from generic initial data 
for a relativistic perfect fluid with $\gamma \alt 1.01$.

\section{Summary and Discussions}
The final fate of the spherically symmetric 
gravitational collapse
of a perfect fluid from time-symmetric initial data
has been investigated numerically.
The $\gamma$-law equation of state
with an adiabatic energy condition was considered.
If $\gamma-1$ is small and the initial density distribution is
not so compact, the collapse of a relativistic fluid 
results in a central,
shell-focusing naked singularity.
The initial data from which a naked singularity occurs
is not zero-measure and therefore sufficiently generic
as long as only time-symmetric and 
spherically symmetric initial data are considered.  
The final fate of a relativistic fluid 
is not a stable star
but either a black hole or a naked singularity.
This is because the total internal energy of
a highly relativistic fluid dominates
the gravitational energy
and therefore the fluid cannot be bound.
We define the critical adiabatic index $\gamma_c$ as an upper
limit of $\gamma$ such that
the collapse of highly relativistic fluid with $\gamma$
can result in the naked singularity formation from generic, 
time-symmetric and spherically symmetric initial data.
The present numerical study shows $\gamma_c\simeq 1.01$. 

If we consider the collapse of an unrelativistic fluid,
the usual picture of the Newtonian gravity is valid.
The collapse of an unrelativistic fluid with $\gamma < 4/3$
does not end up with a stable star, while,
for $\gamma > 4/3$, a stable star is possible.
The final fate of an unrelativistic fluid with $\gamma > 4/3$
is basically determined by energetics.
The numerical code used here 
cannot say whether the final fate of the collapse
of an unrelativistic fluid is a black hole
or a naked singularity because it needs an extremely large
dynamical range. 

Here we compare the results obtained here on the highly relativistic
fluid with the results under the assumption of 
self-similarity by Ori and Piran~\cite{op198719881990}.
The equations of state of both analyses
are approximately common, and hence the only difference will be
the genericity of initial data.  
The value $\gamma_c \simeq 1.01$ which we have obtained here
agrees with the value $\gamma_c\simeq 1.0105$
by Ori and Piran~\cite{op198719881990} within a numerical accuracy.
Note that the initial data prepared here
are time symmetric while 
those of Ori and Piran are imploding due to the
assumption of self-similarity.
We should emphasize that, even with $\gamma<\gamma_c$,
the collapse can result in either a naked singularity 
or a black hole and that it depends on the choice of initial data.

There is a question about whether the results obtained here
support the violation of cosmic censorship.
Is it possible that the adiabatic index $\gamma$ 
(if the adiabatic condition is a good approximation) 
for high-density matter is as
small as $\gamma\alt 1.01$ ?
We do not know the reason why the equation of state becomes so 
soft for relativistically compressed matter 
although we know the adiabatic index $\gamma$ may 
become very small
and even less than unity in some unrelativistic density range~\cite{st1983}.
For example, the radiation fluid, which is generally considered as a good 
approximation for relativistic matter $e\gg 1$,
is given by $e =\infty$ and $\gamma=4/3$.
This might be strong evidence for the validity of cosmic censorship.
However, we are currently not sure of the equation of state 
for highly condensed matter and hence
it remains an open question.

Since a space-time singularity breaks down the smoothness of
physical quantities, numerical simulations cannot give
a rigorous answer about the validity of cosmic censorship.
Nevertheless numerical simulations give physically 
important possibilities
about the maximum density which we can observe
in principle in gravitational collapse, i.e.,
outside an event horizon.
From this point of view, our results show that
we can observe high-energy physics in gravitational collapse 
if the equation of state
of the high-density matter is rather soft.
If it is, the gravitational collapse might be a good
laboratory to obtain clues about high-energy physics.
 
\acknowledgements
I thank W. Israel, T. Nakamura, 
K. Nakao, M. Shibata, H. Iguchi, and K. Omukai
for helpful discussions.
I am grateful to H. Sato
for his continuous encouragements
and helpful discussions.
This work was supported by Grant-in-Aid for 
Scientific Research No.9204
from the Japanese Ministry of Education, Science, Sports
and Culture.

\appendix
\section*{Divergent behavior at the center in the LTB solution}
Here we derive the divergent behavior of a central naked 
singularity
in the marginally bound dust collapse,
which is described by the LTB solution, 
both on the synchronous comoving slice on which the 
naked singularity occurs
and on the earliest null ray which emanates from the central
naked singularity.
For dust, the total energy density coincides with the
rest-mass density identically, i.e., $\rho=\rho_0$.
In synchronous comoving coordinates, the LTB solution of 
marginally bound collapse is given as
\begin{eqnarray}
  ds^2 &=& -dt^2+B^2 dr^2+R^2 d\Omega^2, \\
  R(t,r) &=& \left(\frac{9F}{4}\right)^{1/3}(t_0(r)-t)^{2/3},
  \label{eq:tbr}\\
  B(t,r) &=& R^{\prime}
  \label{eq:tbb}\\
  \rho &=& \frac{1}{8\pi}\frac{F^{\prime}}{R^2 R^{\prime}},
  \label{eq:tbdensity}\\
  t_0(r) &=& \frac{2}{3\sqrt{F}}r^{3/2},
  \label{eq:tbt0}
\end{eqnarray}
where the prime denotes the derivative with respect to $r$ 
and we set $r=R$ at $t=0$ using the rescaling freedom of $r$
in the last equation,
in deriving the last equation.
$F(r)$ is an arbitrary function, a half of which is
the conserved Misner-Sharp mass $m(r)$. 
At $t=t_0(r)$, a singularity occurs at a mass shell 
labeled by $r$.
Here we set $t=t_0(0)$ and assume analytic and generic initial 
data at $t=0$, i.e.,
\begin{equation}
  \rho(t=0,r)=\rho_0+\rho_2 r^2 +\cdots,
  \label{eq:tbinitdensity}
\end{equation}
where we assume that the density is a decreasing function of $r$ and
therefore $\rho_2<0$.
Then, from Eq. (\ref{eq:tbdensity}),
\begin{equation}
  F(r)=F_3 r^3+F_5 r^5 + \cdots,
  \label{eq:tbf}
\end{equation}
where $F_3>0$ and $F_5<0$.
Then, from Eqs. (\ref{eq:tbr}), (\ref{eq:tbb}), and (\ref{eq:tbt0}) 
the following behavior is easily derived at $t=t_0(0)$ for
sufficiently small $r$:
\begin{eqnarray}
  R&\propto&r^{7/3},
  \label{eq:tbcentralr}\\
  R^{\prime}&\propto& r^{4/3}, 
  \label{eq:tbcentralrprime}\\
  F^{\prime}&\propto& r^2.
\end{eqnarray}
From Eq. (\ref{eq:tbdensity}), we find that
\begin{equation}
  \rho\propto r^{-4}.
  \label{eq:tbcentraldensity}
\end{equation}
From Eqs. (\ref{eq:tbf}), 
(\ref{eq:tbcentralr}), and (\ref{eq:tbcentraldensity}),
we conclude that, at $t=t_0(0)$,
\begin{eqnarray}
  \rho &\propto& R^{-12/7}, \\
  \frac{m}{R} &\propto& R^{2/7}.
\end{eqnarray} 
on the synchronous comoving slice $t=t_0(0)$.
This behavior around the naked singularity 
is the same for the nonmarginally bound collapse.
The blow up of the central density is given from Eqs. 
(\ref{eq:tbr}), (\ref{eq:tbdensity}), and (\ref{eq:tbt0}) as
\begin{equation}
  \rho(t,r=0)\propto [t_0(0)-t]^{-2},
\end{equation}
while
\begin{equation}
  \rho(t,r>0)\propto [t_0(r)-t]^{-1}.
\end{equation}

Then we determine the exponents on the earliest outgoing null geodesic
from the central naked singularity
which results from the marginally bound dust collapse. 
On this null geodesic, $R$ and $R^{\prime}$ are 
(see, e.g. Ref.~\cite{jd1993})
\begin{eqnarray}
  R&\propto&r^{7/3}, \\
  R^{\prime}&\propto& r^{4/3}, 
\end{eqnarray}
for $\rho(t=0,r)$ as seen in Fig. (\ref{eq:tbinitdensity}).
Therefore, we conclude that
\begin{eqnarray}
  \rho &\propto& R^{-12/7}, \\
  \frac{m}{R} &\propto& R^{2/7}.
\end{eqnarray}
This behavior seen around the central naked singularity 
is also the same for nonmarginally bound collapse.

\newpage

\begin{table}
  \begin{center}
  \caption{$\tilde{e}_{c}=10^2$}
  \label{tb:hr}
    \begin{tabular}{l|ccccccr} 
        & $ \gamma-1=0$ & $10^{-5}$ & $10^{-4}$ & $10^{-3}$ & $10^{-2}$
      & $10^{-1}$ & $1$
      \\ \hline
      $R_s/M=10$   & B & B & B & B & X & Y & Y \\
          $ 50 $   & N & N & N & N & E & Y & Y \\
          $ 100$   & N & N & N & N & E & Y & Y \\
          $ 1000$  & N & N & N & E & E & Y & Y \\
    \end{tabular}
  \end{center}
\end{table}
\begin{table}
  \begin{center}
  \caption{$\tilde{e}_{c}=10^{2}$}
  \label{tb:hrde}
    \begin{tabular}{l|ccccccr} 
        & $ \gamma-1=2\times 10^{-3}$ & $4\times 10^{-3}$ & $6\times
      10^{-3}$ & $8 \times 10^{-3}$ & $10^{-2}$ & $2\times 10^{-2}$  
      \\ \hline
      $R_s/M=10$   & B & B & BE& X & X & X \\
          $  20$   & N & N & N & N & N & BE\\
          $  30$   & N & N & N & N & N & E \\
          $  40$   & N & N & N & N & E & E \\
          $  50$   & N & N & N & E & E & E \\
          $  60$   & N & N & N & E & E & E \\
          $  70$   & N & N & E & E & E & E \\
          $  80$   & N & N & E & E & E & E \\
          $  90$   & N & N & E & E & E & E \\
          $ 100$   & N & E & E & E & E & E \\
    \end{tabular}
  \end{center}
\end{table}
\begin{table}
  \begin{center}
  \caption{$\tilde{e}_{c}=1$}
  \label{tb:mr}
    \begin{tabular}{l|ccccccr} 
        & $ \gamma-1=0$ & $10^{-5}$ & $10^{-4}$ & $10^{-3}$ & $10^{-2}$
      & $10^{-1}$ & $1$
      \\ \hline
      $R_s/M=10$   & B & B & B & B & BE& E & E \\
          $ 50 $   & N & N & N & N & N & E & E \\
          $ 100$   & N & N & N & N & E & E & E \\
          $ 1000$  & N & N & N & E & E & E & E \\
    \end{tabular}
  \end{center}
\end{table}
\begin{table}
  \begin{center}
  \caption{$\tilde{e}_{c}=10^{-2}$}
  \label{tb:nr}
    \begin{tabular}{l|ccccccr} 
        & $ \gamma-1=0$ & $10^{-3}$ & $10^{-2}$ & $10^{-1}$ & $1/3$ & $2/3$
      & $1$
      \\ \hline
      $R_s/M=10$   & B & B & B & B & B & B & B \\
          $  50$   & N & N & N & N & B & SE& SE\\
          $ 100$   & N & N & N & N & E & SE& SE\\
          $ 1000$  & N & N & N & E & E & E & E \\
    \end{tabular}
  \end{center}
\end{table}
\newpage 


\vskip 0.3in
\centerline{FIGURE CAPTION}
\vskip 0.05in

\newcounter{fignum}
\begin{list}{Fig.\arabic{fignum}.}{\usecounter{fignum}}

\item
Naked-singular model in which $\gamma-1=10^{-4}$, 
$\tilde{e}_{c}=10^2$, and $R_{s}=100M$.
The slicing is by the Misner-Sharp and Hernandez-Misner codes.
The ordinate is the proper time of a comoving observer,
and the abscissa is the circumferential radius.
The Misner-Sharp slicing is $t/M=0, 100, 200, 300, 400, 500, 
600, 700, 707$
and the Hernandez-Misner slicing is 
$u/M=0,100, 200, 300, 400, 500, 600, 700, 728$.
The Hernandez-Misner slicing is a set of outgoing null geodesics.
We stopped the calculation in both codes
when we detected a central singularity. 
Locations of some fluid elements are marked. 
\item 
Evolution of the rest-mass density $\rho_0$ in the Misner-Sharp code.
The density distribution in the central region at $t=707M$ 
becomes so steep that we call it a singularity.
\item
Evolution of $m/R$, the ratio of the Misner-Sharp mass to the
circumferential radius, in the Misner-Sharp code.
\item 
Evolution of $(dR/dt)_{ONG}$, $dR/dt$ along outgoing null geodesics,
in the Misner-Sharp code.
\item 
Evolution of $\rho_0$ in the Hernandez-Misner code.
The density distribution in the central region at $u=728M$ 
becomes so steep that we call it a singularity. 
\item 
Evolution of $m/R$ in the Hernandez-Misner code.
\item
Evolution of the lapse function $e^{\psi}$
in the Hernandez-Misner code.
\item 
Blow up of the central rest-mass density $\rho_{0c}$.
The simulation was repeated with various radial grid resolutions.
Each curve is labeled by the number of spatial zones used. 
\item 
Black-hole model in which 
$\gamma-1=10^{-4}$, $\tilde{e}_{c}=10^2$, and $R_{s}=10M$.
The Misner-Sharp slicing is $t/M=0, 10.0, 20.0, 22.3$.
On the last slice, a singularity was detected.
The Hernandez-Misner slicing is 
$u/M=0,10.0, 20.0, 30.0, 40.0, 50.0, 60.0, 70.0, 72.3$.
On the last slice, an event horizon was detected.
The limit curve is the event horizon.
\item 
Evolution of $\rho_0$ in the Misner-Sharp code.
The density distribution in the central region at $t=22.3M$ 
becomes so steep that we call it a singularity.
\item
Evolution of $m/R$ in the Misner-Sharp code.
\item
Evolution of $(dR/dt)_{ONG}$ 
in the Misner-Sharp code.
\item
Evolution of $\rho_0$ in the Hernandez-Misner code.
The density distribution in the central region 
remains not so steep even at the event horizon.
\item 
Evolution of $m/R$ in the Hernandez-Misner code.
\item 
Evolution of $e^{\psi}$ in the Hernandez-Misner code.
The $e^{\psi}$ converges to zero as $u$ increases.
\item 
Evolution of $\rho_{0c}$.
The simulation was repeated with various radial grid resolutions.
Each curve is labeled by the number of spatial zones used. 
\item 
Stable-star model
in which $\gamma-1=2/3$, $\tilde{e}_{c}=10^{-2}$, and $R_{s}=100M$.
The Misner-Sharp slicing is $t/M=0$, $2.00\times 10^4$, 
$4.00\times10^4$, 
$6.00\times 10^4$, $8.00\times 10^4$, $1.00\times 10^5$, $1.20\times 10^5$,
$1.22\times 10^5$.
The Hernandez-Misner slicing is 
$u/M=0$, $2.00\times 10^4$, $4.00\times 10^4$, $6.00\times 10^4$, 
$8.00\times 10^4$, $1.00\times 10^5$, $1.18\times10^5$.
\item   
Evolution of $\rho_0$ in the Misner-Sharp code.
The density distribution of the core 
settles down after $t \simeq 6.00\times 10^4$.
\item
Evolution of $m/R$ in the Misner-Sharp code.
\item 
Evolution of $(dR/dt)_{ONG}$ in the Misner-Sharp code.
\item
Evolution of $\rho_{0c}$.
The simulation was repeated with various radial grid resolutions.
Each curve is labeled by the number of spatial zones used. 
\end{list}

\newpage
\end{document}